\documentclass[11pt]{article}
\usepackage{amsfonts}
\usepackage{amssymb}
\usepackage{graphicx}
\usepackage[bottom]{footmisc}

\begin{document}
\title{
{Springer series in Physics}\\
  {\small to be published.}\\
  \ \\
{\bf Ferroelectricity and Charge Ordering in Quasi One-Dimensional
Organic Conductors}
 }
\author{{\huge Serguei Brazovskii}
\ \\ \ \\
 LPTMS-CNRS, UMR8626, Univ. Paris Sud, bat 100,  ORSAY CEDEX, F-91405\\
and \\
L.D. Landau Institute for Theoretical Physics, Moscow, Russia\\
 \ \\
 \texttt{brazov@lptms.u-psud.fr}}

\bigskip\bigskip
\date{7 November 2006}

\maketitle

\begin{abstract}
The family of molecular conductors TMTTF/TMTSF-X demonstrates almost
all known electronic phases in parallel with a set of weak
structural modifications of \lq anion ordering\rq\ and mysterious
\lq structureless\rq\ transitions. Only in early 2000s their nature
became elucidated by discoveries of a huge anomaly in the dielectric
permittivity and by the NMR evidences for the charge ordering
(disproportionation). These observations have been interpreted as
the never expected ferroelectric transition. The phenomenon unifies
a variety of different concepts and observations in quite unusual
aspects or conjunctions: ferroelectricity of good conductors,
structural instability towards the Mott-Hubbard state, Wigner
crystallization in a dense electronic system, the ordered $4K_F$
charge density wave, richness of physics of solitons, interplay of
structural and electronic symmetries. The corresponding theory of
the "combined Mott-Hubbard state" deals with orthogonal
contributions to the Umklapp scattering of electrons coming from the
two symmetry breaking effects: the built-in nonequivalence of bonds
and the spontaneous nonequivalence of sites. The state gives rise to
several types of solitons, all of them showing in experiments. On
this basis we can interpret the complex of existing experiments, and
suggest future ones, such as optical absorption and
photoconductivity, combined ferroelectric resonance and the phonon
anti-resonance, plasma frequency reduction.
\\

\bigskip\bigskip \noindent{\bf keywords}:
Organic conductors, ferroelectricity, charge ordering, charge
disproportionation, solitons, optics, conductivity, permittivity.
\\

Viewgraphs:\\
http://ipnweb.in2p3.fr/~lptms/membres/brazov/seminars.html

\end{abstract}

\newpage
\begin{small}
{\tableofcontents}
\end{small}

\newpage

\section{Introduction. History and Events. \label{sec:intro}}
 The discovery of first organic superconductors (Bechgaard, Jerome, Ribault, et al - 1979/80,
see\footnote{We start the literature review with proceedings of most
important scientific conferences and reviews collections
\cite{ICSM-82,yamada,ISCOM-03,ISCOM-05,ICSM-02,ICSM-04,ecrys-02,ecrys-05,dekker,schegol,reidel}.}
\cite{jerome:94,jerome:04}) gave rise to a number of directions
which none could foresee in advance. A great deal of experiments and
theoretical speculations were done already within first years.
Still a firework of phases was so amusingly rich that it feeds intensive research till our
days. The twentieth anniversary was marked by a new discovery of Ferroelectricity and
Charge Ordering/Disproportionation whose consequences are the subject of this review.

We shall address only quasi one-dimensional (1D) systems, and among them mostly the family
of the Bechgaard-Fabre salts $\mathrm{(TMTCF)_2X}$, with extensions to related materials.
Typically, the selenium subfamily $\mathrm{(TMTSF)_2X}$ was a basis for research on the
metallic phase: particularly the superconductivity and diverse spectacular magnetic
oscillations. The sulphur subfamily $\mathrm{(TMTTF)_2X}$ was suitable for effects of
electronic correlations, particularly the Mott-Hubbard dielectrization. The studies were
mostly concentrated upon electronic phase transitions, which take place between 1K and 20K,
and on fashionable "non Fermi liquid" aspects of the normal state at higher T. Actually
this "normal" part of the phase diagram was filled, for most of compositions X, with tiny
structural transitions of "anion orderings" observed for all non-centrosymmetric
counterions X (\cite{moret}, and \cite{anions} for a review). Intentionally or not, these
features were not appreciated either in experiments on electronic properties, or in
theories - except notes by Emery \cite{emery:83} and a systematic approach of Yakovenko and
the author \cite{BY:phys-let,BY:jetp}.

The most powerful sleeping bomb was hidden in mysterious "structureless
transitions"registered in sulphur subfamily for centrosymmetric anions (X=Br, AsF$_6$,
etc.) at higher temperatures of about 100-250K (\cite{lawersanne,coulon,javadi}, and
\cite{coulon:iscom-03} for a recent review). Unexplained, unattended by the community and
finally forgotten, these transitions have been waiting for a revenge since mid 80's. Only
it 2000's  their nature was identified, almost by chance, as something never expected: the
ferroelectricity (FE) \cite{monceau:01}, or sometimes antiferroelectricity (AFE), see
reviews \cite{nad:02,nad:06}. Moreover, the ferroelectricity happened to be a consequence
of a hidden charge ordering/disproportionation (CO/CD) \cite{brown}%
\footnote{The terms Charge Ordering (CO), Charge Disproportionation (CD) and Charge
Localization have been utilized in the literature to identify the described phenomena. Here
we shall typically use a more easily pronounced version of charge ordering, and reserve the
term charge disproportionation for the same effect to underline nonequal molecular charges.
The charge ordering transition temperature will be called $T_{\mathrm{CO}}$; its other
notations in the literature were typically $T_{0}$ or $T_{CD}$.}.
 The whole high temperature region of the phase diagram happened to be symmetry-broken.
More history excursions can be found in
\cite{coulon:iscom-03,brown:iscom-03,seo:iscom-03}, see also Appendix \ref{sec:history}.

The phenomenon is gaining ground, as it is being found in more and more compounds
\cite{coulon:iscom-03,batail,ravy:iscom-03}. Also the charge ordering  is registered now in layered organic
materials, see \cite{fukuyama,takahashi:ecrys-05} for reviews,
accompanied by the anomalously high dielectric permittivity \cite{nad-bedt}, also signs of
the ferroelectricity have been indicated \cite{matsui:03}. The ferroelectricity was
observed also in dielectric mixed-stacks organic compounds showing the neutral-ionic
transition \cite{rennes}. The energy scale (up to $300K$ in terms of temperatures and $\sim
1000K$ in terms of energy gaps) of the charge ordering  effect is so high that it allows to
anticipate its hidden presence also in the metallic subfamily $\mathrm{(TMTSF)_2X}$; the
first direct sign has been brought by the case of $\mathrm{(TMTSF)_2FSO_3}$
\cite{takahashi:ecrys-05}. At least the effect must determine the properties of many
materials where its presence has already been confirmed.

\begin{figure}[htb]
\centering\includegraphics*[width=0.7\textwidth]{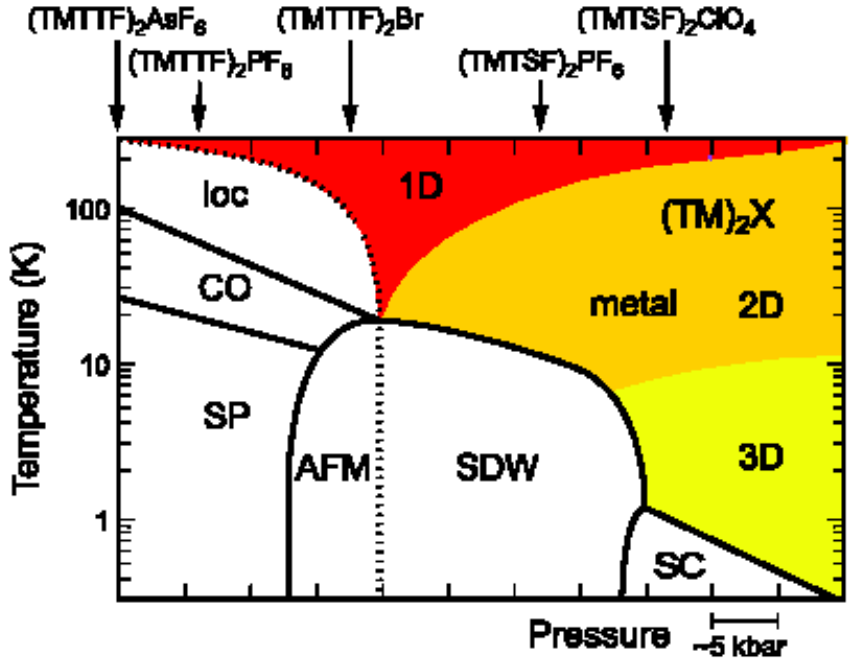} \caption{Schematic phase
diagram of the $\mathrm{(TMTCF)_2X}$ family. SC- superconductivity;  AFM -
antiferromagnet - i.e. commensurate SDW; SP - Spin-Peierls; CO- charge ordering, loc -
charge localization (CO pre-transitional effect); 1D,2D,3D - dimensional regimes. After
Dressel, Dumm et al.} \label{fig:phase-diag}
\end{figure}

How could it happen, that such a dominant effect, which is so clear/obvious today, was
missed, with so many misleading consequences? Actually there was more than enough of
warnings through decades. In the prehistoric epoch\footnote{A unique review connecting the
science of organic metals throughout all decades is offered in \cite{jerome:04}}, it was
the structural 4K$_{F}$ anomaly, for that time this discovery could also be classified as a
never expected one (\cite{pouget-4kf,kagoshima,NMP-TCNQ}; \cite{emery-4kf} for a
theory,\cite{4kf} for a review). Its impact upon electronic properties has been shown via
the strong effect of the 4K$_{F}$+2K$_{F}$ lock-in \cite{jerome:78}, unfortunately this
spectacular observation was washed out from memory. The coexisting 4K$_{F}$ and 2K$_{F}$
charge density waves (CDWs) with the exact trimerization were observed in NMP-TCNQ
\cite{NMP-TCNQ}; remarkably, the T dependent conductivity resembles the one of
$\mathrm{(TMTTF)_2X}$. The sequence of transitions (spontaneous dimerization followed by
the spin-Peierls (SP) tetramerization) in MEM-TCNQ \cite{mem} can be compared directly with
today's cases of $\mathrm{(TMTTF)_2PF_{6}}$ and $\mathrm{(TMTTF)_2ReO_{4}}$, see Sect.
\ref{sec:solitons-combined}.

Soon after the discovery of organic superconductors, a very rich information has been
accumulated on subtle structural transitions of the anion orderings (see the review
\cite{anions}), but these "gifts of the magi" were not exploited as much as they deserve.
As an exemption, in mid 80's the interplay of electronic and structural properties was
emphasized in the earlier construction of the universal phase diagram
\cite{BY:phys-let,BY:jetp}; its milestones were just the cases of the intersite charge
ordering  in $\mathrm{(TMTTF)_2SCN}$ and the interchain charge disproportionation in
$\mathrm{(TMTSF)_2ClO_4}$ (in contemporary language).

\begin{figure}[htb]
\centering
\includegraphics[width=0.7\textwidth]{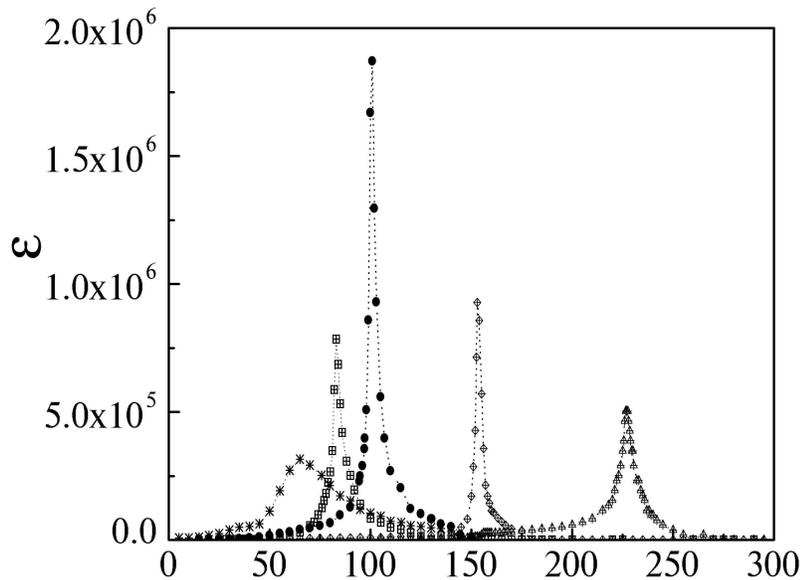}
\caption{Linear plots of the temperature dependence of the real part of the dielectric
permittivity $\varepsilon^{\prime}$ at 100 kHz for $\mathrm{X=PF_6}$ (stars),
$\mathrm{BF_4}$ (squares), $\mathrm{AsF_6}$ (circles), $\mathrm{SbF_6}$ (diamonds),
$\mathrm{ReO_4}$ (triangles). After \cite{nad:06}, see also a figure in \cite{nad:02}b.}
\label{fig:eps-lin-mps}
\end{figure}

In retrospect of the structureless transitions, the events started to accelerate in 1997
when a theoretical work by Seo and Fukuyama \cite{seo:97} indicated that the AFM state may
be accompanied by the spontaneous inequivalence of on-cite occupations - the charge
disproportionation. Next year, the NMR experiments by Hiraki and Kanoda \cite{kanoda} have
allowed to register the charge disproportionation  in a new compound
$\mathrm{(DI-DCNQI)_2Ag}$, with understanding of observation of the Wigner crystallization
or equivalently the condensation of the $4K_F$ anomaly\footnote{The charge ordering  had
been already found \cite{ilakovac:95} in substituted perylene salts $\mathrm{(TMP)_2X}$
($\mathrm{X=PF_6, AsF_6}$), which was left unattended, may be because of the accent upon
effects of disorder. The charge ordering  was identified via combined observations of the
$\mathrm{4K_F}$ CDW by the X-ray and of the doubling of sites by the $\mathrm{C^{13}}$
NMR.}. The SDW ordering was present indeed, but at a temperature $T_{SDW}$ which is more
than one order of magnitude lower than the charge ordering  one $T_{\mathrm{CO}}$. One year
later, in 1999, Nad and Monceau \cite{nad:2000}, pursuing other goals of physics of sliding
incommensurate SDWs, have faced the structureless transitions. Their methodic has allowed
to precisely pinpoint the transition temperatures and to show that the already pronounced
anomaly in microwave response \cite{javadi} becomes unsustainably high in the low frequency
experiments of \cite{nad:2000}. This strong anomaly and the precise localization of its
temperature have attracted the attention of the Brown group, and in 2000 they registered by
the NMR the spontaneous inequivalence of sites since its very onset at $T_{\mathrm{CO}}$.
It took another year to finally resolve the mystery of the structureless transitions in
2001 \cite{monceau:01}. The remarkable improvement in the experimental  techniques has
allowed to obtain anomalies as sharp as shown in Fig.\ref{fig:eps-lin-mps}. This, together
with the suggested theory \cite{monceau:01,brazov:ICSM-02}, left no doubt that we are
dealing with a least expected phase transition to the ferroelectric state. This second
order transition manifests itself in the giant anomaly of the dielectric permittivity
$\varepsilon(T)$, Fig.\ref{fig:eps-lin-mps}, which perfectly fits the Landau-Curie-Weiss
law, Fig.\ref{fig:1/eps(X)} below.

The ferroelectricity is ultimately coupled with the charge ordering,
which is followed by a fast formation, or a steep increase, of the
conductivity gap $\Delta$. There is no sign of a spin gap formation
\cite{chi} or of a spin ordering down to the ten times lower scale
of $T_e$. Hence we are dealing with a surprising ferroelectric
version of the Mott-Hubbard state which usually was associated
rather with magnetic orderings. The anomalous diverging
polarizability is coming from the electronic system, even if ions
are very important to choose and stabilize the long range $3D$
order. The ferroelectric transition in $\mathrm{(TMTTF)_{2}X}$ is a
particular, bright manifestation of a more general phenomenon of
charge ordering, which now becomes recognized as a common feature of
organic and some other conductors
\cite{fukuyama,alloul,takahashi:ecrys-05}.

 This rich history tells us about the necessity for reconciliation of different
branches
of Synthetic Metals which have been almost split for two decades%
\footnote{The conference ICSM-82 \cite{ICSM-82} was both the summit in science of
Synthetic Metals and the last meeting where its all directions were present
simultaneously and in mutual recognition.}. %
 Indeed, the major success in finding the ferroelectric anomaly was due to precise low frequency
methods for the dielectric permittivity $\varepsilon$ \cite{monceau:96}:  designed for
pinned CDWs in inorganic chain conductors \cite{Nad:93}, the methods were applied also to
SDWs in organic conductors \cite{nad-sdw}, and finally to the structureless transitions
\cite{nad:2000}. On the theory side, the author's approach of the Combined Mott-Hubbard
state \cite{monceau:01,brazov:iscom-01,brazov:ICSM-02,brazov:ecrys-02} has been derived
from a similar experience \cite{ABx,rice} in a model of conducting polymers.

\section{Hierarchy of phases in quasi 1D organic conductors. \label{sec:phases}}
Low dimensional electronic systems serve as a workshop on both particular and  general
problems of strong correlations, degenerate collective ground states, their symmetry and
topological features. The richest opportunities have been opened by the family of first
quasi $1D$ organic superconductors: the Bechgaard - Fabre salts $\mathrm{(TMTSF)_{2}X}$,
$\mathrm{(TMTTF)_{2}X}$. These compounds demonstrate, at low temperatures $T$, transitions
to almost all known electronic phases, see \cite{jerome:94,jerome:04} and
Fig.\ref{fig:phase-diag}. These are: a normal metal, a regime with the Mott-Hubbard charge
localization (the (para)magnetic insulator -MI in classification of
\cite{BY:phys-let,BY:jetp}), a spin-Peierls (i.e. a charge density wave of bonds
dimerization in a framework of strong electronic repulsion), a spin density wave (SDW), an
antiferromagnet (AFM) (i.e. the doubly commensurate SDW), a SDW induced by high magnetic
fields (FISDW), and finally a superconductivity (SC) (of still a suggestive nature
\cite{triplet}, see also Appendix \ref{sec:history}). Temperatures of electronic
transitions $T=T_{e}$ range within 1K for the SC, 10-15K for SDWs, 10-20K for the
spin-Peierls phase, see Fig.\ref{fig:phase-diag}.

\subsection{Structural transitions of the anion ordering. \label{sec:anions}}

\begin{figure}[htb]
\centering
\includegraphics[width=0.4\textwidth]{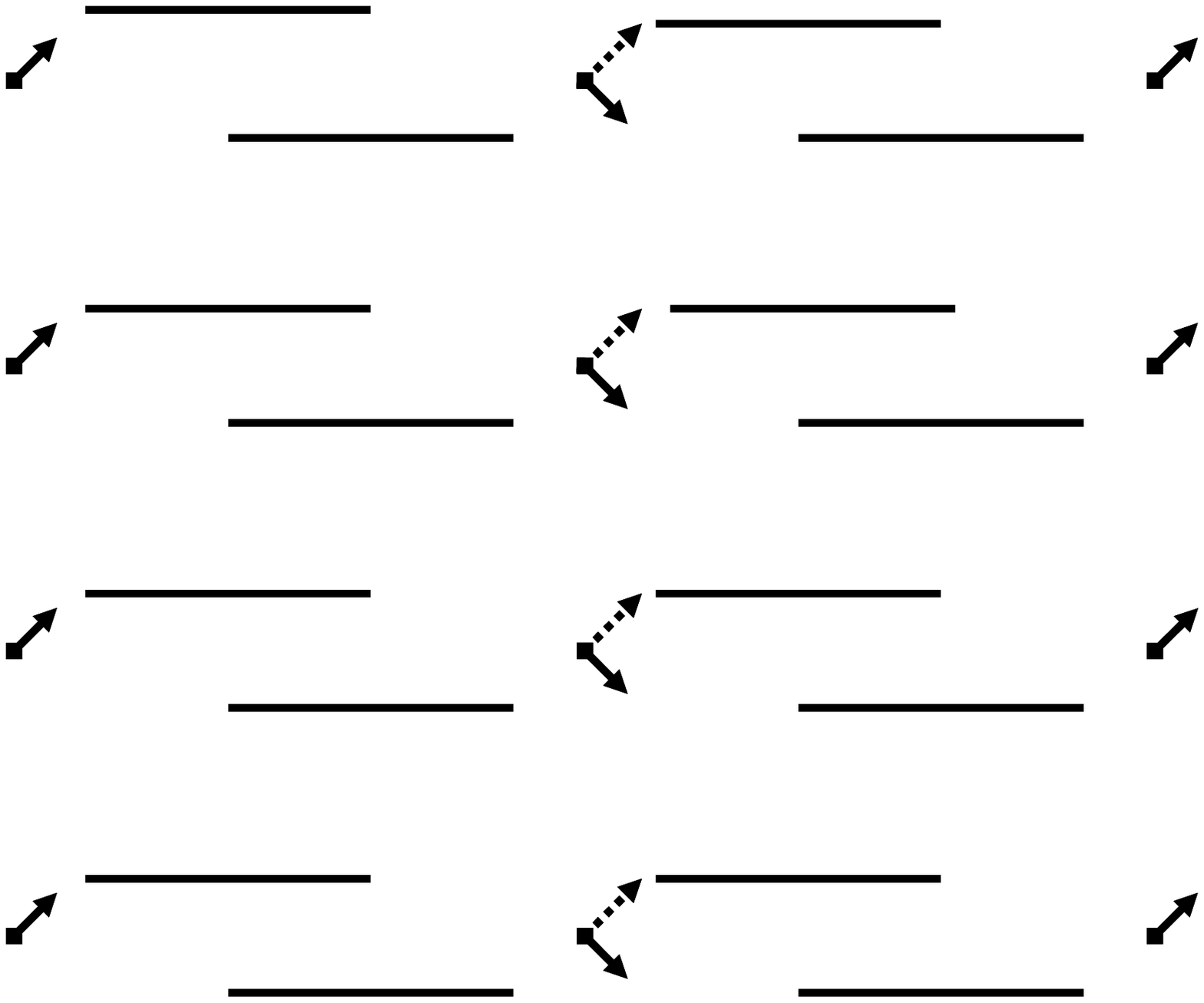}
\hskip2cm
\includegraphics[width=0.4\textwidth]{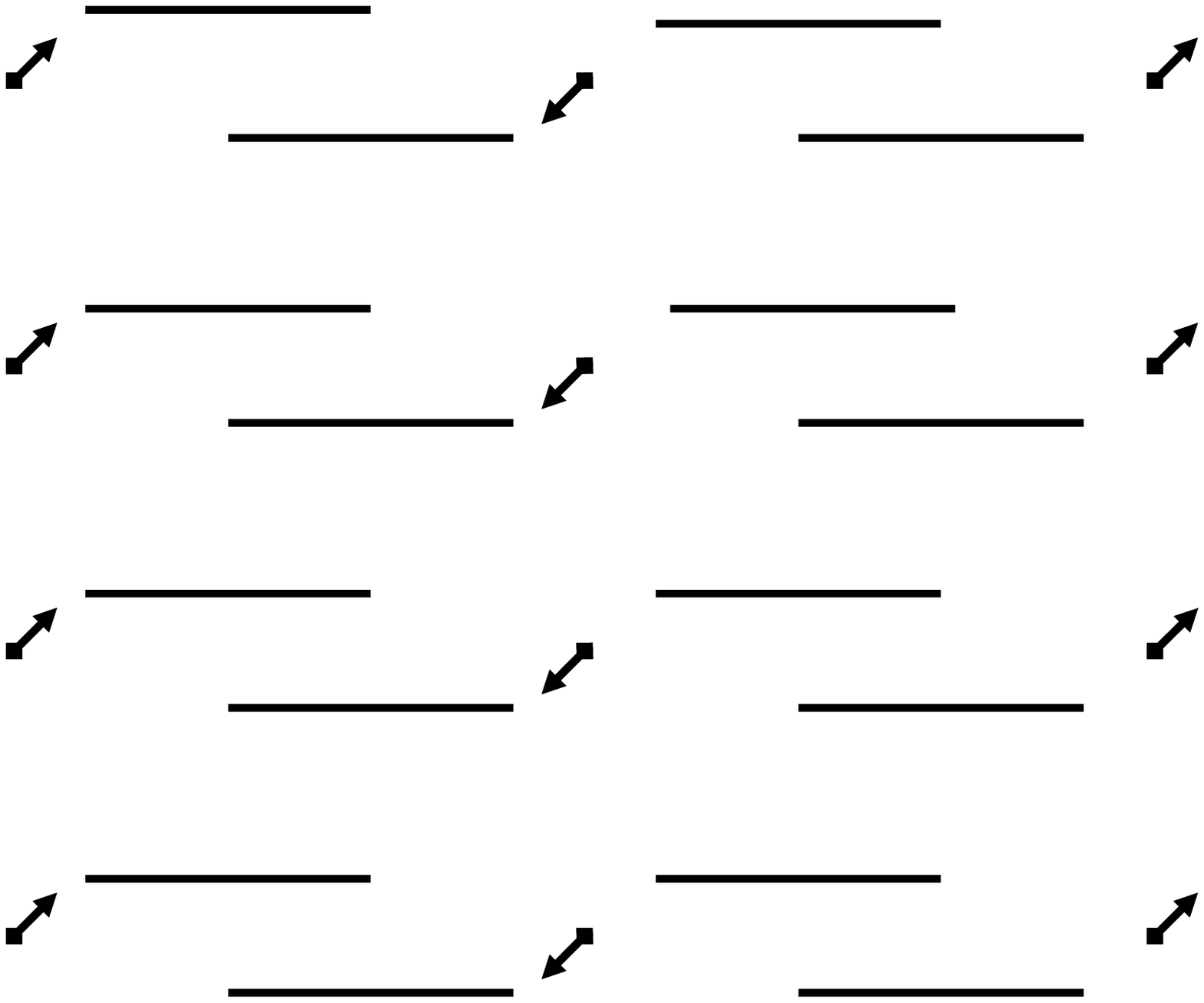}
\caption{Structure and selected instabilities of $\mathrm{(TMTCF)_2X}$ compounds.
Horizontal bars show the molecules TMTCF, side view; the intermolecular  spacings
alternate. Filled squares show the anions X; arrows show displacements/orientations of
ions at various AOs.  Left panel: ${\mathbf q}_2$ structure of the AFE type in
$\mathrm{(TMTTF)_2 SCN}$ (all solid arrows); if the dotted arrows are chosen, then we
arrive at the model for the ferroelectric \lq{structureless}\rq\, phase. In both cases
the molecules become nonequivalent, hence the intersite CD. Right panel: ${\mathbf q}_3$
structure of the relaxed $\mathrm{(TMTSF)_2ClO_4}$. Molecular sites are kept equivalent
within the chain, but chains environment alternate, hence the interchain CD. Other,
tetramerized structures are shown in \cite{anions}.}
 \label{fig:tmttf}
\end{figure}

There is a set of several weakly different structural types due to the anion ordering (AO)
\cite{anions}, which are fine arrangements of chains of singly charged counterions
$\mathrm{X}$. The temperatures $T_{\mathrm{AO}}$ of these weak structural transitions are
usually about 100K. The AOs are characterized, as any superstructure, by their wave numbers
${\mathbf q}=(q_{\parallel},q_{\perp})\neq0$, where $q_{\parallel}$ and $q_{\perp}$ are the
components in the stack and the interstack directions. The AOs were always observed for
non-centrosymmetric ions (typically tetrahedral X=ClO$_4$,ReO$_4$,BF$_4$, etc., except for
the special case of the linear ion X=SCN). The orientational ordering was supposed to be a
leading mechanism, see \cite{bruinsma:83} for a theory,  with positional displacements
(arrows in Fig.\ref{fig:tmttf}) being its consequences only. But there is also a universal
mechanism of the Earnshow instability \cite{earnshow}, inherent to all structures of
classical charged particles, which can trigger independently the displacive component in
materials with both centrosymmetric and non-centrosymmetric anions. (See more in the
Appendix \ref{sec:earnshow}.)

Already within the unperturbed crystal structure at $T>T{_\mathrm{CO}}$, the tiny
dimerization of bonds by anions $\mathrm{X}$ in $\mathrm{(TMTCF)_2X}$ provokes the
dielectrization \cite{zagreb,emery:82,BY:jetp}. Regular alternation of ions, positioned
against every second intermolecular spacing, dimerizes the intermolecular distances. It
doubles the on-stack unit cell, hence changes the mean electronic occupation from $1/2$ per
molecule to $1$ per dimer. The bond alternation gives rise \cite{zagreb} to the relatively
small Umklapp scattering $U_{b}$, which opens (according to Dzyaloshinskii and Larkin
\cite{DL}, Luther and Emery \cite{LE,emery:79}, see Sects.
\ref{sec:theory1},\ref{sec:theory2}) the route to the Mott-Hubbard insulator. (See
\cite{brazov:iscom-01} and \cite{brazov:ICSM-04} for a history introduction and Appendix
\ref{sec:history} for some quotations).

\subsection{Charge ordering transitions. \label{sec:co}}
At similar or even higher range, $T=T_{\mathrm{CO}}\approx 100\div 200K$, also other
unidentified transitions were observed sometimes in the $\mathrm{TMTTF}$ subfamily. Their
signatures were seen in conductivity \cite{lawersanne}, microwave permittivity
\cite{javadi}, in thermopower \cite{coulon}. But there was no trace of any lattice effects,
hence the title \lq{structureless}\rq\ transitions was assigned.  Recently their mysterious
nature has been elucidated by discoveries of the huge anomaly in the dielectric
permittivity $\varepsilon$ at $T_{\mathrm{CO}}$ \cite{monceau:01,nad:02} and of the charge
disproportionation seen by the NMR \cite{brown} at $T<T_{\mathrm{CO}}$. It is still an
experimental challenge to refine the structure; a scheme for its reconstruction is outlined
in this section.

The ferroelectric anomaly signifies that the term \lq structureless\rq\, is not quite
correct: this is not an isomorphic modification. While the Brave lattice is based on the
same unit cell indeed, the full space symmetry is changed by loosing the inversion
center, which is the necessary condition for the ferroelectric state.

The NMR experiments \cite{brown} have clearly detected the appearance at the
$T_{\mathrm{CO}}$ of the site nonequivalence as a sign of the CO/CD. Recently the charge
disproportionation was also confirmed by means of the molecular spectroscopy
\cite{dumm:ecrys-05}, see Fig. \ref{fig:dumm} below. While bonds are usually dimerised
already within the basic structure, see Fig. \ref{fig:tmttf}, the molecules stay equivalent
above $T_{\mathrm{CO}}$, the latter symmetry being lifted by the charge ordering
transition. At presence of the inequivalence of bonds, the additional inequivalence of
sites lifts the inversion symmetry, hence the allowance for the chain's polarization
leading to the ferroelectricity.

The confidence for this identification comes from a good fortune, that the 3D pattern of
the charge ordering  appears in two, AFE and FE, forms. Both types AFE and FE are the
same paramagnetic insulators (the MI phase of \cite{BY:phys-let,BY:jetp}), which is
shown by comparison of electric and magnetic measurements.  The NMR brought the same
observation \cite{brown} both for the structureless transitions and for the particular
AFE phase in $\mathrm{(TMTTF)_{2}SCN}$; also the permittivities at high frequencies and
temperatures are very similar.  The structure of the $\mathrm{(TMTTF)_{2}SCN}$ has been
already known as a very rare type of $\vec{q}_{2}=(0,1/2,1/2)$. The site inequivalence
was already identified by the structural studies, see \cite{anions} and Fig.
\ref{fig:tmttf}. The structure was completely refined; the anions displacements are
unilateral for each column while undulating among them.

So, what is the difference between the $\vec{q}_{2}$ case and the \lq{structureless}\rq\,
transitions? The $\vec{q}_{2}$ structure showed already the common polarization along a
single stack, $q_{\parallel}=0$, but alternating in perpendicular directions,
$\vec{q}_{\bot}\ne 0$, Fig. \ref{fig:tmttf}; it gives, as we can interpret today, the AFE
ordering. For the \lq{structureless}\rq\, transition all displacements must be identical
$\vec{q}=0$ among the chains/stacks, thus leading fortunately to the ferroelectric state.
We arrive at the quite sound conjecture that the ferroelectric 'structureless' state
corresponds to the scheme of Fig. \ref{fig:tmttf} where we should choose the dotted arrows
instead of solid ones. Without the advantage of having the structural information
\cite{anions} on the $\mathrm{(TMTTF)_{2}SCN}$, our understanding of the nature of the
ferroelectric compounds would be more speculative. This case can be viewed as a
corner-stone, or Rosetta stone, in a sense that it belongs to two classes of anion ordering
and charge ordering systems and transferrers a complementary information.

\begin{figure}[htb]
\centering\includegraphics*[width=0.7\textwidth]{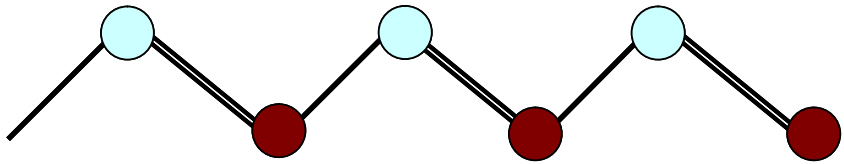} \caption{The scheme of a
conjugated polymer of the (AB)$_x$ type, realized recently as the modified polyacetylene
\cite{mpa}. Dark and light circles correspond to chemically inequivalent monomers
(actually the carbon atoms with different ligands). Single and double connections show
the expected spontaneous dimerization of bonds. Curiously, this model takes just the
form of the hypothetical design by W. Little for a polymeric superconductor (D. Jerome,
introduction to ISCOM-03 \cite{ISCOM-03}), which suggestion gave birth to the whole
science of Synthetic Metals.} \label{fig:ABx}
\end{figure}

In any case, FE or AFE, the polar displacement gives rise to the joint effect of the
built-in and the spontaneous contributions to the dimerization, due to alternations of both
bonds and sites\footnote{Both contributions can be of the built-in type in the particular
case of the $\mathrm{(TMTSF)_{0.5}(TMTTF)_{0.5}}$ mixture \cite{0.5}, where the alternation
of S/T-containing molecules provides the site dimerization. Oppositely, both coexisting
dimerizations may appear spontaneously, but these two independent symmetry breakings
require for two successive second order phase transitions or for the first order one. The
later option may be relevant to observations \cite{ravy:iscom-03,nad-bedt}.}. None of these
two types of dimerization changes the unit cell of the zigzag stack which basically
contains two molecules, hence $q_{\parallel}=0$ for the charge ordering  wave vector. But
their sequence lifts the mirror (glide plane) and then the inversion symmetries which must
lead to the on-stack electric polarization. This interference resembles the orthogonal
mixing in the "combined Peierls state" \cite{ABx} in conjugated polymers of the $(AB)_{x}$
type, Fig. \ref{fig:ABx}. Here also the bond dimerization is the $q_\parallel=0$
transition, just because of the backbone zigzag structure: what looks for electrons as a
period doubling, is structurally the lifting of the glide plane symmetry.

\subsection{Overlapping and coexistence of phases.}
The \lq{structureless}\rq\, displacive instability and the usual \lq{orientational}\rq\ AOs
can be independent. This is supported by finding of a sequence of the ferroelectric and the
anion ordering transitions in $\mathrm{(TMTTF)_{2}ReO_{4}}$ \cite{nad:02}. We will show in
Sect. \ref{sec:solitons-combined} that this true present from the Nature gives an access to
compound solitons, as to intriguing events of the spin-charge reconfinement. At the same
time, the case of $\mathrm{(TMTTF)_{2}SCN}$, where $q_\parallel=0$ while $q_\bot\ne0$,
presents the corner-stone belonging to both types of AOs and charge ordering transitions.

Notice finally that the $q_\parallel=0$ structure is not ultimately the on-chain charge
ordering. The so-called $\vec{q}_3=(0,1/2,0)$ type of the anion ordering \cite{anions},
observed in the relaxed phase of the $\mathrm{(TMTSF)_2ClO_4}$, shows the interchain
redistribution of charges. This is a kind of the "incommensurability transition" (with an
unusually reversed order in comparison with the conventional lock-in to the commensurable
state) which may be considered \cite{BY:phys-let,BY:jetp} as a prerequisite of the
superconductivity.

\subsection{Electronic mechanism of the charge ordering. \label{sec:gs}}

To get an idea of the driving force behind the hidden
\lq{structureless}\rq\,  transition we shall follow the example of
the ''combined Peierls state'' \cite{SB:84,matv} developed for
polymers like the modified polyacetylene. It describes a joint
effect of built-in and spontaneous contributions to the
dimerization, hence to the electronic gap. Since this concept was
formulated \cite{SB:81}, it has been widely used in studies of
conducting polymers gaining a clear experimental support in diverse
observations, especially in optics and ESR, see  \cite{trans-cis}.
Particularly relevant is the experimentally recent \cite{mpa}case of
the ''orthogonal mixing'' relevant to polymers of the $(AB)_{x}$
type, Fig. \ref{fig:ABx}. Here the built-in gap $\Delta_{s}$ comes
from the site dimerization due to the $AB$ alternation, while the
spontaneous contribution to the total gap $\Delta$ comes from the
dimerization of bonds $\Delta_{b}$, like in the generic Peierls
effect. Importantly, the two gaps add in quadrature:
$\Delta=\sqrt{\Delta_{in}^{2}+\Delta_{ex}^{2}}$ - the amplitudes of
electron's scattering, Fig. \ref{fig:umklaps} - right, are shifted
by $\pi/2$.

For the present case of the charge ordering, the Peierls effect is
substituted by the Mott-Hubbard one. Also the built-in and the
spontaneous effects are interchanged: the built-in one comes from
inequivalence of bonds while the spontaneous one comes from non
equivalence of sites. The charge gap $\Delta =\Delta (U)$ appears as
a consequence of both contributions to the Umklapp scattering, and
it is a function of the total amplitude
$U=\sqrt{U_{s}^{2}+U_{b}^{2}}$. (But the observable gap $\Delta(U)$
is not additive in quadratures any more.)

\begin{figure}[htb]
\includegraphics[width=0.7\textwidth]{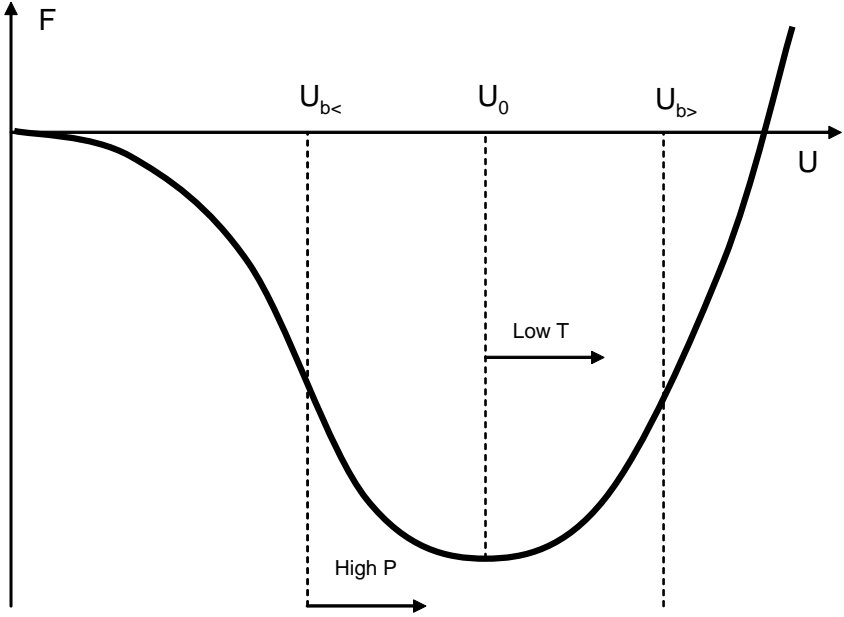}\\
\caption{Ground state energy $F(U)$ as a function of the total magnitude
$U=\sqrt{U_{s}^{2}+U_{b}^{2}}$ of the commensurability (Umklapp) potential. Its minimum
at $U_0$ is reachable if $U_b<U_0$ corresponding to the position $U_{b<}$; then the
spontaneous $U_s$ is created to augment $U_b$ to $U_0$. It does not happen, if $U_b>U_0$
corresponding to the position $U_{b>}$, then there is no transition. The arrows show
that these two regimes can interchange by either changing of $U_0$ by temperature or of
$U_b$ by pressure.}
  \label{fig:energy}
\end{figure}

The electronic energy $F_{e}$ depends only on the charge gap
$\Delta$, which is a function of only the total
$U=\sqrt{U_{s}^{2}+U_{b}^{2}}$. The energy of lattice distortions
$F_l$ depends only on the spontaneous site component $U_{s}$:
$F_{l}=$ $(K/2)U_{s}^{2}$, where $K$ is an elastic constant. Thus
the total energy can be written in terms of the total $U$:
\[
F(U)=F_{e}(U)+1/2KU^{2}-1/2KU_{b}^{2}\; , \; U\geq U_{b}
\]
The ground state is determined by its minimum over $U$, but unusually at the constraint
$U\geq U_{b}$. We can encounter three possibilities for the energy function $F(U)$,
Fig.\ref{fig:energy}.

a. It has no minimum except $U=0$: no charge ordering  at any condition;

b. It has a minimum at some value $U=U_{0}<U_{b}$: charge ordering  is possible but not
reached yet;

c. It has a minimum at some value $U=U_{0}>U_{b}$, which now determines the ground state.
Since the value $U_{0}$ increases with decreasing temperature, there will be a phase
transition at $U_{0}(T)=U_{b}$ provided that $U_{0}(0)>U_{b}$.

The phase transition in the regime (c) can be reversed if $U_b$ is made to increase
passing above $U_0$, then the charge ordering  disappears. This is what seems to happen
typically in experiments under pressure \cite{nagasawa:ISCOM-03,pressure}.

How does this minimum at $U_0$ appear actually? In principle, the
electron energy $F_{e}(U)<0$ is always gained by opening the gap
$\Delta$ which reduces the zero point fluctuations. But to overcome
the energy loss $\sim U_s^2$ from lattice deformations and charge
disproportionation, we need $F_{e}(U)\sim -U^{2-\zeta}$ with
$\zeta\geq 0$. For conventional $2K_F$ CDWs it is always the case
with $F_{e}(U)\sim -U^{2}\ln E_{F}/U$ corresponding to the limit
$\zeta\rightarrow 0$. For our case of the $4K_{F}$ modulations the
condition is reached only at large enough interactions
\cite{emery-4kf} with the marginal case $\zeta=0$ corresponding to
the so-called Luther-Emery line \cite{LE,emery:79}, see more in
Sect.\ref{sec:theory1}. We can view it also as a general criterium
of the $4K_{F}$ instability \cite{emery-4kf}.

\subsection{Electric polarization and Ferroelectricity. \label{sec:fe}}
In principle, there are three contributions to the electric polarizability:%
\newline 1. \textit{{Intergap electronic polarizability}} is regular at $T_{\mathrm{CO}}$:
$\varepsilon_{\Delta}\sim{\omega_{p}^{2}}/{\Delta^{2}}$, where
$\omega_{p}$ is the plasma frequency. At the transition, where
 $\Delta(T\approx T_{\mathrm{CO}})$ is still well below its low $T$ value $\Delta(0)$,
$\varepsilon_{\Delta}$ can be as large as $\sim10^3$ which
corresponds indeed to  the background upon which the anomaly at
$T_{\mathrm{CO}}$ is developed. The value of the gap well
corresponds to the lower $\varepsilon_{\Delta}\approx 100$ as it was
estimated already in \cite{javadi}.
\newline 2.
\textit{{Ion displacements}} could already lead to the macroscopic polarization, like in
usual ferroelectrics, but taken alone they cannot explain the observed giant magnitude of
the effect. Taking the  typical parameters \cite{anions} of AOs (recall the SCN case as the
corner-stone), the ionic displacive contribution may be estimated (see Appendix
\ref{sec:epsilon-ions}) as
$\varepsilon_{i}\sim10^{1}{T_{\mathrm{CO}}}/|T-T_{\mathrm{CO}}|$, which is by $10^{-3}$
below a typical experimental value
$\varepsilon\approx2.5\times10^{4}T_{\mathrm{CO}}/(T-T_{\mathrm{CO}})$
\cite{monceau:01}.%
\newline 3.
\textit{Collective electronic contribution} can be estimated roughly
\cite{monceau:01}, see Appendix \ref{sec:epsilon-ions}, as a product
of the above two
$\varepsilon_{el}\sim\varepsilon_{i}\varepsilon_{\Delta}$, which
provides both the correct $T$ dependence and the right order of
magnitude of the effect. The anomalous diverging polarizability is
coming from the electron subsystem, even if the instability is
triggered by the ions, whose role is to stabilize the long range
$3D$ ferroelectric order, and to discriminate between FE, AFE, or
more complex patterns \cite{ravy:iscom-03}, see more in the Appendix
\ref{sec:e-i}.

\section{Electronic properties. \label{sec:exp1}}

Here we shall give a short summary of experimental results on electronic properties of
$\mathrm{(TMTTF)_2X}$ compounds \footnote{Experimental plots of this chapter are the
results by Monceau, Nad, et al, see \cite{nad:02,nad:06} for a broad collection.}.

\subsection{Permittivity.}

\begin{figure}[htb]
\centering\includegraphics*[width=0.7\textwidth]{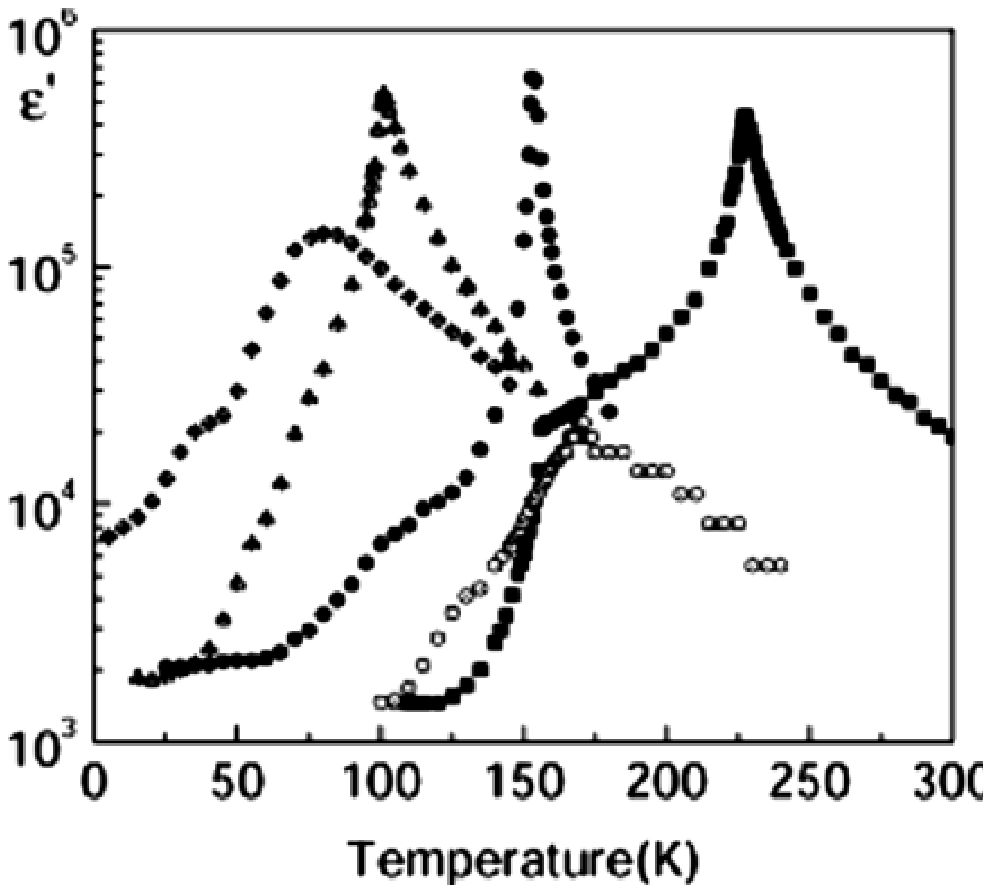}
\caption{$\log\varepsilon^\prime$ vs $T$ at f=1MHz for FE cases $\mathrm{X=PF_6}$
($\Diamond$), $AsF_6$ ($\triangle $), $SbF_6$ ($\bullet$), $ReO_4$ ($\blacksquare$) and
for the AFE $\mathrm{X=SCN}$ ($\odot$). After \cite{nad:02}.} \label{fig:eps-log}
\end{figure}

\begin{figure}[htb]
\centering\includegraphics*[width=0.7\textwidth]{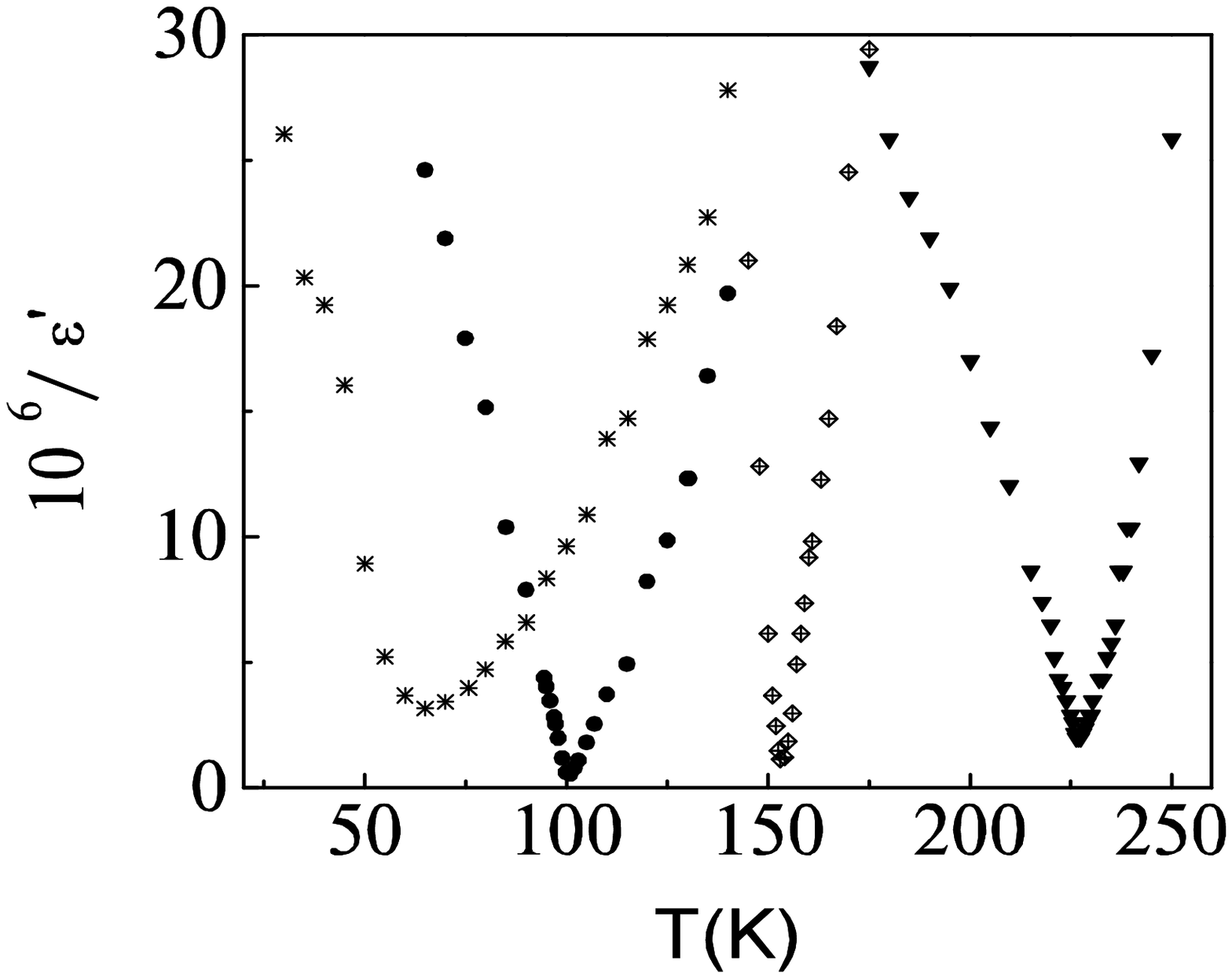} \caption{Temperature
dependence of the inverse of the real part of the dielectric permittivity
$\varepsilon^{\prime}$ of $\mathrm{(TMTTF)_2X}$ with $\mathrm{X= PF_6, AsF_6, SbF_6,
ReO_4}$ (in the order from left to the right) at the frequency of $100Hz$. After
\cite{monceau:01,nad:02}.} \label{fig:1/eps(X)}
\end{figure}

\begin{figure}[htb]
\centering
\includegraphics*[width=0.7\textwidth,angle=90]{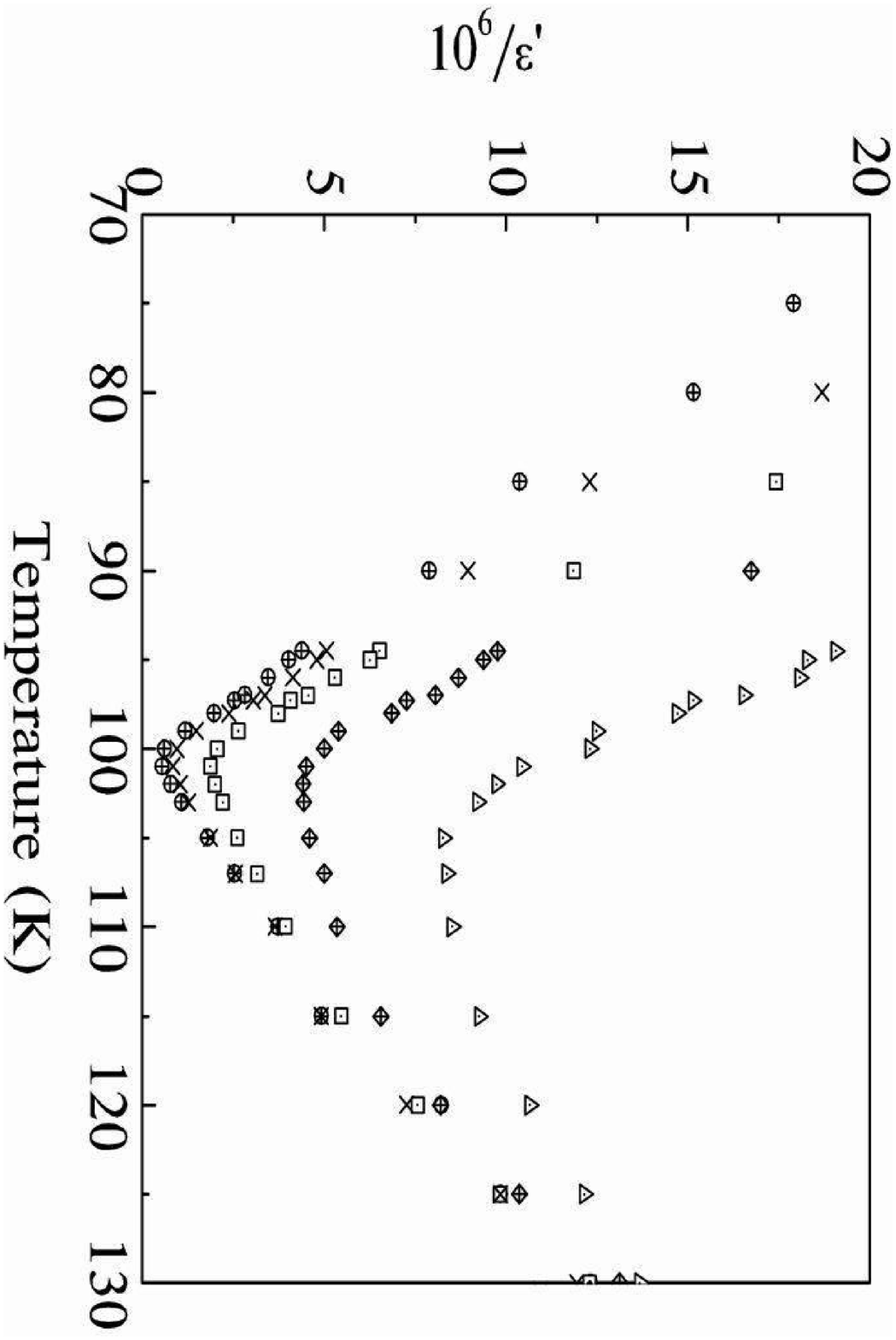}
\caption{Plots of $10^6/\varepsilon^\prime$(T) of $\mathrm{(TMTTF)_2AsF_6}$ at
10000,3000,1000,300,100 KHz. Nad, Monceau et al, unpublished.} \label{fig:disp}
\end{figure}

Typical plots of the temperature dependence of the dielectric permittivity
$\varepsilon$\footnote{We imply everywhere, unless specified, the real part of
$\varepsilon$: $\Re\varepsilon=\varepsilon^\prime$} at
Figs.\ref{fig:eps-lin-mps},\ref{fig:eps-log}, demonstrate very sharp (even in the $\log$
scale of Fig. \ref{fig:eps-log}!) ferroelectric peaks. Most of cases show the purely
mono-domain "initial" ferroelectric permittivity, ideally fitting the Curie law
$\varepsilon\sim|T-T_{\mathrm{CO}}|^{-1}$, Fig. \ref{fig:1/eps(X)} - even with the right
ratio $=2$ of the Landau theory for inclinations at $T\gtrless T_{\mathrm{CO}}$
\cite{monceau:01} \footnote{A relatively rounded anomaly in the PF$_{6}$ case of Fig.
\ref{fig:1/eps(X)} correlates with its stronger frequency dispersion, Fig.
\ref{fig:disp}: it can be a pinning of ferroelectric domain walls, in other words a
hidden hysteresis. For an extended information on frequency dispersion, see
\cite{nad-tau,nad:ecrys-05} and the review \cite{nad:06}.}.
 In general, we observe the frequency dependent depolarization of the mono domain ferroelectricity,
instead of the more usual long time hysteresis of the re-polarization. Still abundant
normal carriers screen the ferroelectric polarization at the surface, which eliminates the
need for the domain structure. Low-temperature studies are necessary to find the remnant
polarization. Radiational damage or other disorder will also help to pin the domain walls
and freeze the polarization, which should give rise to the conventional hysteresis curve.

The AFE case of $\mathrm{X=SCN}$ in  Fig. \ref{fig:eps-log} shows a smooth maximum of
$\varepsilon$, rather than a divergent peak, as it should be. Nevertheless, the high $T$
slopes of all curves for $\varepsilon(T)$ look very similar. It tells us that the
ferroelectric state is gradually developing already within the high $T>T_{\mathrm{CO}}$ 1D
regime, before the 3D interactions discriminate between the in-phase FE and the
out-of-phase AFE orderings. Recall that, contrary to these low frequencies, in the
microwave range the X=SCN compound  showed \cite{javadi} the strongest response with
respect to ferroelectric cases; hence the purely 1D regime of the ferroelectricity was
recovered. This may also be the key to the AFE/FE choice. Indeed, for highly polarizable
units containing the already polar ion SCN, the dominating Coulomb forces will always lead
to an AFE structure.

The $\mathrm{X=ReO_4}$ case shows at 150K a subsequent first order
anion ordering transition of tetramerization. Here $\varepsilon$
drops down (Fig. \ref{fig:eps-log}, see also \cite{nad:02,nad:06}
for details) which might be caused (via the factor
$\omega_p^2/\Delta^2$) by the increase of the gap $\Delta$ as it is
seen at the conductivity plot in Fig. \ref{fig:ReO4+SCN}. The case
of $\mathrm{X=PF_6}$ shows the spin-Peierls transition at T=19K.
While being symmetrically equivalent to the anion ordering in
$\mathrm{X=ReO_4}$ case, it is of the second order, so that in
$\varepsilon(T)$ it shows up only as a shoulder.

\subsection{Conductivity.}
\begin{figure}[htb]
\centering
\includegraphics*[width=0.7\textwidth]{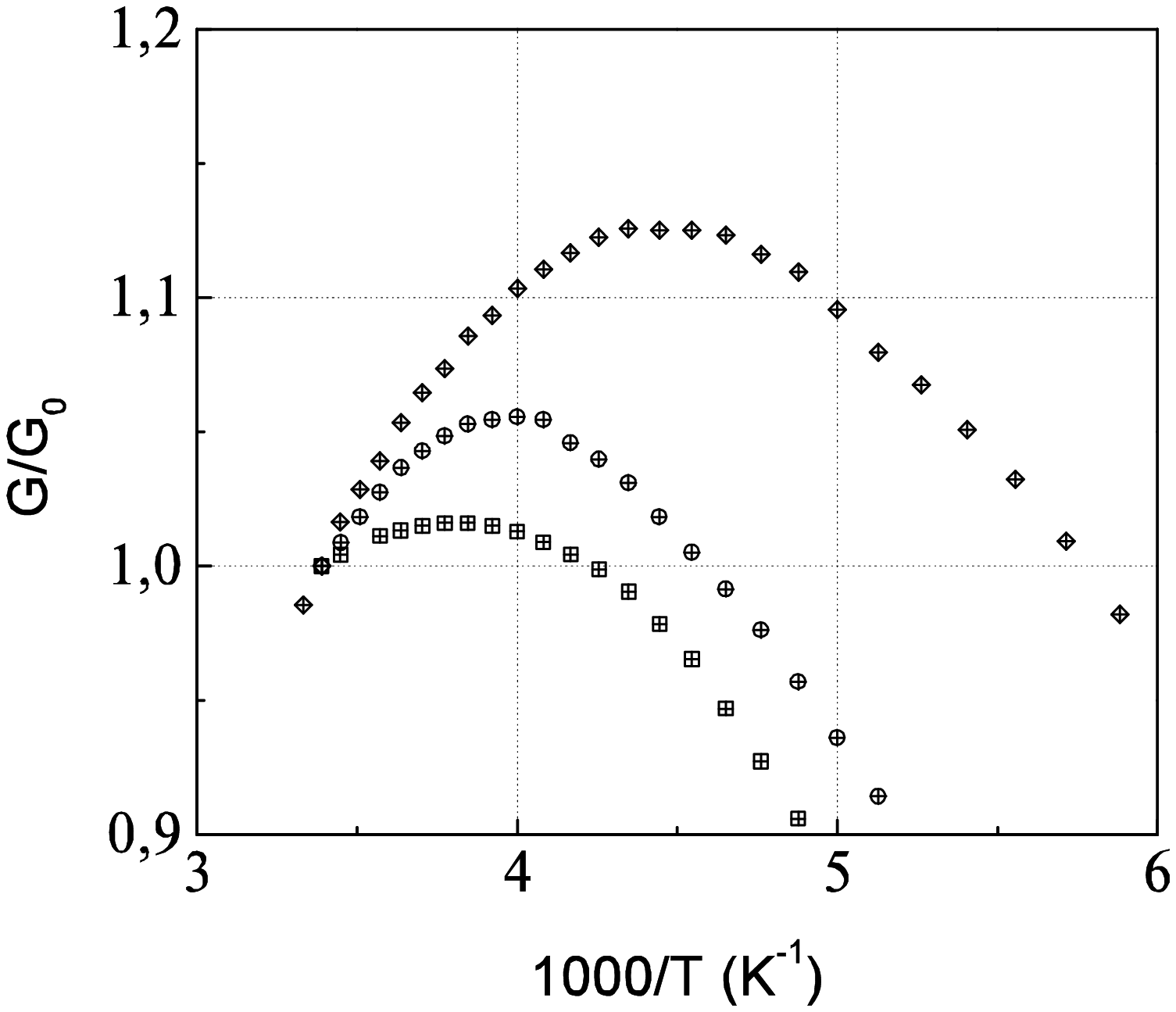}
\caption{Plots for the normalized conductivities $G/G_0$ ($G_0$ is the RT value) in the
$T\approx T_{\rho}$ region. $\mathrm{X = SbF_6}$ (squares), $\mathrm{AsF_6}$ (circles),
$\mathrm{PF_6}$ (diamonds). The plots $G(T)$ show maxima at $T=T_\rho>T_{\mathrm{CO}}$.
After \cite{nad:02,nad:06}.} \label{fig:cond}
\end{figure}

\begin{figure}[htb]
\centering
\includegraphics*[width=0.7\textwidth]{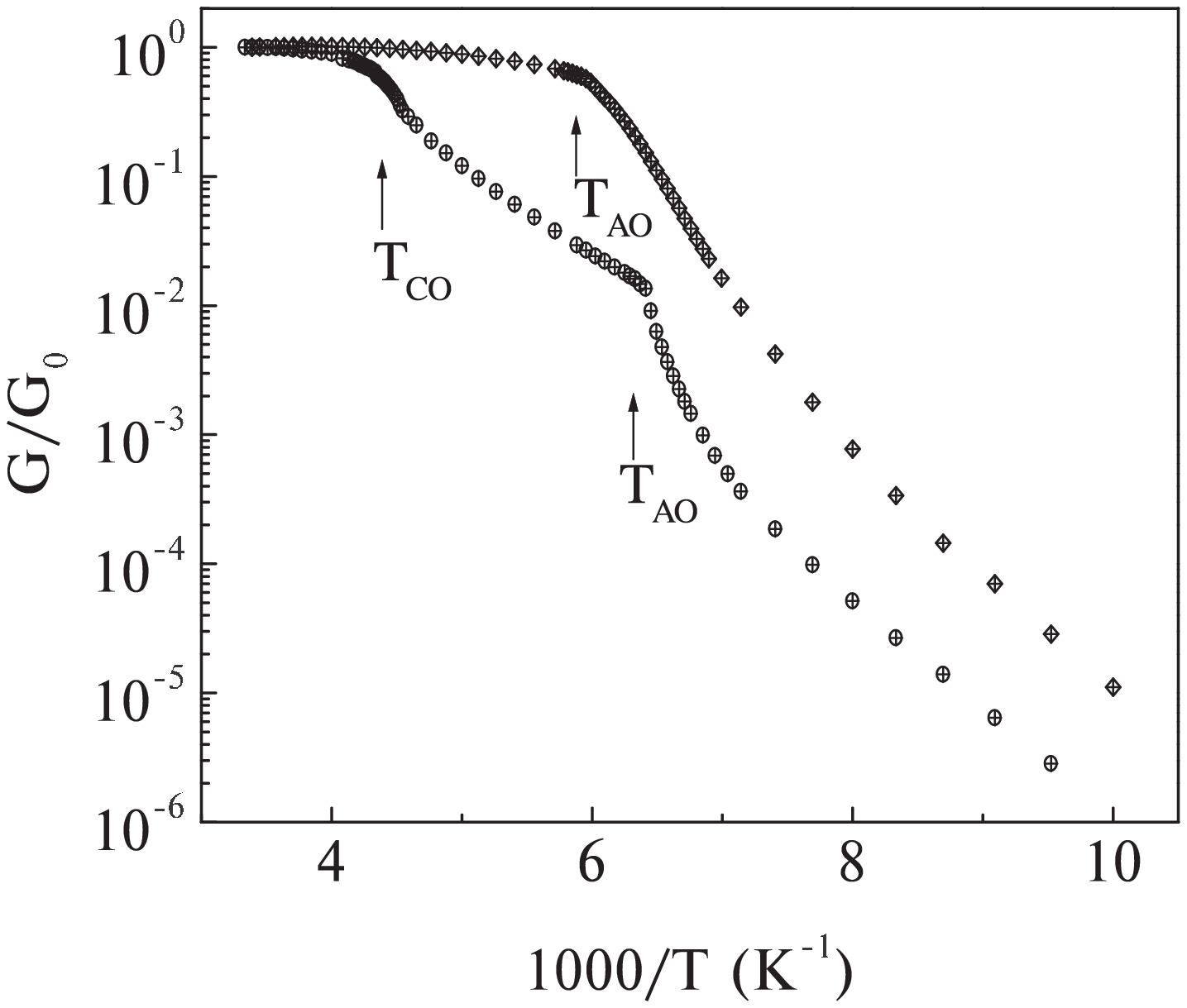}
\caption{Ahrenius plots for normalized conductivities $G/G_0$ in a wide $T$ region.
$\mathrm{X=ReO_4}$ ($\circ $), $\mathrm{X=SCN}$ ($\Diamond$). After
\cite{reo4+scn,brazov:ICSM-02}.} \label{fig:ReO4+SCN}
\end{figure}

Fig. \ref{fig:cond} gives several examples of the conductivity G(T) within the high $T$
region around $T_{\rho}$. Fig. \ref{fig:ReO4+SCN} shows, for selected examples, the
conductivity G(T) within a broad range of temperatures (see more cases in
\cite{nad:02,nad:06}). Typical plots correlate with old data \cite{javadi}\footnote{In
reviewed experiments by Nad, Monceau et al, the conductivity was extracted via
$\Im\varepsilon$, within the same experiment as measuring the $Re\varepsilon$.}, but with
more insight available today:

1) Clear examples for conduction by charged spinless particles - holons = solitons: there
are no gaps in spin susceptibility $\chi\approx cnst$ \cite{chi};

2) There is no qualitative difference in G(T) (as well as in NMR ! \cite{brown}) between
FE cases and the AFE one of the SCN: only the on-chain charge ordering  is important,
not the type of the interchain pattern!

3) Coexistence of both types of transitions, CO and AO, in $\mathrm{(TMTSF)_2ReO_4}$; the
subsequent anion ordering increases the conduction gap and opens the spin gap
\cite{spin-gap-reo4}.

4) Gap contribution of the spontaneous site dimerization develops very fast, and soon it
dominates over the bond dimerization gap. The last may not be seen at all -- recall that
the ferroelectric anomaly extends by at least 30K above $T_{\mathrm{CO}}$, Fig.
\ref{fig:1/eps(X)}, signifying the pre-transitional CO, which is able to workout a
pseudogap.

At $T<T_{\mathrm{CO}}$ the charge ordering  adds more to the charge
gap $\Delta$, which is formed now by joint effects of alternations
of bonds and sites. The conductivity $G(T)$ plots in Fig.
\ref{fig:cond} show this change by kinks at $T=T_{\mathrm{CO}}$,
turning down to higher activation energies at low $T$. The steepness
of $G$ just below $T_{\mathrm{CO}}$ reflects the growth of the
charge ordering contribution to the gap, which is expected to be
$\delta\Delta(T)\sim\sqrt{T_{\mathrm{CO}}-T}$, see section
\ref{sec:theory1}; it must correlate with $\varepsilon^{-1}\sim
(T_{\mathrm{CO}}-T)$. Indeed, what may look as the enhanced gap (the
tangent for the Ahrenius plot) near $T_{\mathrm{CO}}$, actually can
be its $T$ dependence expected as $\Delta(T)=\sqrt{\Delta_b^2
+C\Delta_s(0)(1-T/T_{\mathrm{CO}})}$, $C\sim 1$. The differential
plotting of $\Delta(T)=-d\log G/d(1/T)$ would be helpful.

So marginal effect of the built-in bond dimerization opens the route to compounds with
equivalent bonds \cite{kanoda,coulon:iscom-03,batail,heuze:03}.

$\mathrm{(DMtTTF)_2ClO_4}$: Here, the ordered state has been already
identified \cite{ravy:iscom-03} as a complex (incommensurate in the
transverse direction) AFE structure, which develops following a
fascinating pseudo-first order phase transition at T=150K. This
material shows \cite{coulon:iscom-03} the characteristic resistivity
rise below quite typical $T_\rho\approx 120-150K$, which does not
affect, as expected, the susceptibility $\chi$ until the low
$T_{SDW}\approx 10K$ AFM ordering. The charge ordered phase is
waiting to be tested for the NMR splitting and for the $\varepsilon$
anomaly, or at least for traditional signatures of structureless
transitions like the thermopower.

$\mathrm{(EDT-TTF-CONMe_2)_2X}$: A similar situation is expected in this new material
\cite{coulon:iscom-03,batail,heuze:03}, which is also the Mott insulator with the
conductivity gap $\approx 1350K$. Unfortunately, there is a lack of information for this
compound; vaguely it was declared \cite{heuze:03} as a clear case of 4-fold
commensurability effects, following the "no CO/CD" scenario for $\mathrm{(TMTCF)_2X}$
family, see Sec. \ref{sec:where?}.

$\mathrm{(DI-DCNQI)_2Ag}$: The nondimerized compound, where the observation of the
charge ordering  \cite{kanoda} has started the modern trend. Here also the activation
energy value $\approx 490K$ is well above the $T_{\mathrm{CO}}\approx 220K$.

Now we can close the circle to return to the $\mathrm{(TMTTF)_2X}$ series and to guess that
the observed conductivity downturn at $T_\rho>T_{\mathrm{CO}}$ may be mostly due to the
fluctuational pseudogap coming from the charge ordering  proximity. There may be only a
minor contribution from the bond dimerization specific to $\mathrm{(TMTTF)_2X}$.

\section{Ferroelectric Mott-Hubbard ground state. \label{sec:theory1}}
\subsection{Choosing the theory approach. \label{sec:method}}
The earliest theoretical prediction \cite{seo:97} applies literally to a situation expected
in 3D, or at most 2D systems, where the charge ordering  is set up simultaneously and in
ultimate conjunction with the AFM/SDW order. In $\mathrm{(TMTTF)_{2}X}$ type cases, the
pronouncedly $1D$ electronic regime brings its particular character, as well as it allows
for a specifically efficient treatment \cite{monceau:01}, which also happens to be
particularly well suited to describe the ferroelectric transition.

We are using the bosonization procedure, see reviews \cite{emery:79,solyom,gogolin}, which
is most adequate to describe low energy excitations and collective processes. It takes into
account automatically the separation of spins and charges, which is a common feature of our
systems. All information about basic interactions, whatever they are (on-site, neighboring
sites, long range Coulomb, lattice contribution, see Sect. \ref{sec:generic}) is
concentrated in a single parameter ($\gamma=K_\rho$ - in modern notations). This approach
allows to efficiently use the symmetry arguments and classification. It allows to naturally
interpret the solitonic spinless nature of elementary excitation - the thermally activated
charge carriers. Most importantly for our goals, this approach provides a direct access to
the dielectric permittivity. The procedure easily covers also the secondary anion ordering
or spin-Peierls transitions at lower $T$. A rigorous way to describe a correlated 1D
electronic system, bosonization also provides a physically transparent phenomenological
interpretation in terms of fluctuating $4K_{F}$ density wave, i.e. a local Wigner crystal,
subject to a weak commensurability potential. See more arguments in \cite{BY:jetp}, and
some quotations in Appendix \ref{sec:history}.

Among other approaches, recall also the traditional line of RG theories in recent quasi
1D versions \cite{bourbon:iscom-03,giamarchi:iscom-03}. The today's results do not pass
the major test for the charge ordering  (at least it was merely overlooked), hence they
cannot be applied to $\mathrm{(TMTTF)_{2}X}$ type cases, even if a work was deducted to
this purpose \cite{bourbon:iscom-03}.

Recall also numerical studies like exact diagonalization of small clusters with short range
interactions or quantum chemistry methods \cite{mazumdar,poilblanc}. Usually they pass the
test for the charge ordering, but experience difficulties to obtain its ferroelectric type
as the most favorable one. Even in case of a success for the ground state, the access to
polarization and to elementary excitations will not be sufficient, or even possible,
particularly for more sophisticated cases of combined symmetry breakings, Sect.
\ref{sec:solitons-combined}.

\subsection{Ground state and symmetry breaking. \label{sec:symmetry}}
We introduce the charge phase $\varphi$ as for 2K$_{F}$ CDW or SDW modulations
$\sim\cos(\varphi+(x-a/2)\pi/2a)$ - the origin is taken at the inversion center
in-between the two molecules. Here $a$ is the intermolecular distance; recall that
$\pi/4a=K_F$ is the Fermi number in the metallic praphase. Later on we shall need also
the spin counting phase $\theta$: $\varphi$ and $\theta$ together define the CDW order
parameters completely as ${\cal O}_{CDW}\sim\exp(i\varphi)\cos\theta$. Gradients of
these phases $\varphi^{\prime}/\pi$ and $\theta^{\prime}/\pi$ give local concentrations
of the charge and the spin. The energy density (potential and kinetic) of charge
polarizations is
\begin{equation}
\frac{1}{4\pi\gamma}\left\{(\partial_{x}\varphi)^{2}v_{\rho}+
(\partial_{t}\varphi)^{2}/v_{\rho} \right\} \label{H-phi}
\end{equation}
Here $v_{\rho}\sim v_F$ is the charge sound velocity and $\gamma$ is the main control
parameter, which depends on all interactions; its origin is discussed in Sect.
\ref{sec:generic}. ($\gamma$ is the same as $\gamma_{\rho}$ of \cite{BY:phys-let,BY:jetp}
or $K_{\rho}$ of our days.)

\begin{figure}[htb]
\centering\includegraphics*[width=.4\textwidth]{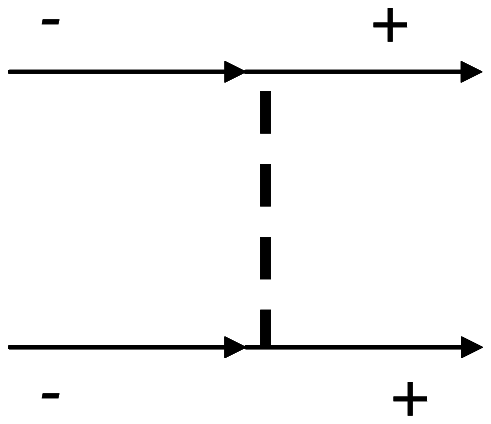} \qquad
\includegraphics*[width=.4\textwidth]{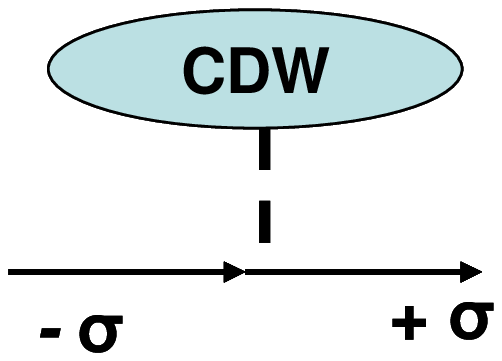}
\caption{Umklapp processes. Left: an electron pair is scattered from $-K_F$ to $+K_F$; the quasi-momentum
deficiency $4K_F$ is absorbed by the periodicity of the dimerised lattice. Right: at presence of the
tetramerization providing the $2K_f$ CDW background, a single electron is scattered from $-K_F$ to $+K_F$
(conserving the spin $\sigma$); the deficiency $2K_F$ is absorbed by the CDW order corresponding to the
tetramerization.} \label{fig:umklaps}
\end{figure}

In addition to (\ref{H-phi}), there is also the commensurability energy $H_U$ coming
from the Umklapp scattering of electrons, Fig. \ref{fig:umklaps}a-left. For our
particular goals it is important to notice several interfering sources for the weak two
fold commensurability, i.e. two contributions to the Umklapp interaction. Their forms
can be derived from the symmetry alone \cite{monceau:01}, as we shall sketch now.

Consider the non dimerized system with $1/2$ electrons per site. It possesses translational
and inversion symmetries $x\rightarrow x+1$ and $x\rightarrow -x $ which corresponds to
phase transformations $\varphi \rightarrow \varphi +\pi/2$ and $\varphi \rightarrow
-\varphi$. The lowest order invariant contribution to the Hamiltonian is $H_4\sim U_{4}\cos
4\varphi$. This is the 4-fold commensurability energy which is usually very small, recall
the conventional CDWs. The reason is not only a smallness of $U_{4}$ which is coming from
Umklapp interaction of $8$ particles, half of them staying high away from the Fermi energy.
In addition, it is renormalized as $\sim \exp [-8<\varphi^{2}>]$, that is it becomes small,
being a product of two large numbers in the negative exponent, see more in Sect.
\ref{sec:where?}.

Consider next the site dimerization which modulates alternatively the onsite energies.
The preserved symmetries are $x\rightarrow x+2$ and $x\rightarrow 1-x$ (reflection with
respect to the molecular cite center). Hence the invariance is required with respect to
$\varphi\rightarrow\varphi+\pi$ and $\varphi \rightarrow\pi/2-\varphi$ and we arrive at
$H_{U}^{s}=-U_{s}\sin 2\varphi$.

Consider finally the bond dimerization. The symmetry $x\rightarrow x+1$ is broken while
the symmetries $x\rightarrow x+2$ and $x\rightarrow -x$ are preserved. The Hamiltonian
must be invariant under the corresponding transformations of $\varphi$:
$\varphi\rightarrow\varphi+\pi$ and $\varphi\rightarrow-\varphi$ and we arrive at
$H_{U}^{b}=-U_{b}\cos 2\varphi$.

At presence of both types of dimerization, the nonlinear Hamiltonian, to be added to
(\ref{H-phi}), becomes

\begin{eqnarray}
H_{U}= -U_b\cos 2\varphi -U_s\sin 2\varphi=-U\cos(2\varphi-2\alpha) \;;\; \\
\nonumber
U=(U_{b}^{2}+ U_{s}^{2})^{1/2} \, ,
\, \tan 2\alpha = U_{s}/ U_{b}
\label{H-U}\end{eqnarray}

The $4k_{F}$ charge modulation will be
\[\rho_{4K_{F}}\sim
\psi_{-}^{2}\psi_{+}^{\dagger 2}+HC\sim \cos (\pi x/a+2\alpha)=(-1)^{x/a}\cos(2\alpha)
\]
where $\psi_\pm$ are one-electron operators near the points $\pm K_F$. The $2K_{F}$ bond
and site CDWs are proportional to
\[
\rho_{2K_{F}}^{b,s}\sim \{\psi_{-}\psi_{+}^{\dagger}\mp h.c.\} \sim\cos\theta\{\cos
,\sin\} (\pi x/2a+\alpha)
\]
Their mean values are zero, because of the spin factor which vanishes in average
$<\cos\theta >=0$ until the spin gap is established below the spin-Peierls temperature,
see Sect. \ref{sec:solitons-combined}.

The $U_s$ comes from the  electronic charge ordering  and ionic
displacements coupled by long range 3D Coulomb and structural
interaction, which are well described by the mean field approach.
But the electronic degrees of freedom must be treated exactly at a
given $U_s$. The renormalization, due to quantum fluctuations of
phases, leaves the angle $\alpha$ invariant, but it reduces $U$ in
(\ref{H-U}) down to $U^{\ast}\sim \Delta^{2}/\hbar v_{F}$
($U^{\ast}\neq 0$ only at $\gamma <1 $); it determines the gap
$\Delta\sim U^{1/(2-2\gamma)}$. (In scaling relations, we imply
units $E_F$ for $\Delta$ and $E_F/a$ for $U$.) The spontaneous
charge ordering  $U_{s}\neq 0$ requires that $\gamma<1/2$, far
enough from $\gamma =1$ for noninteracting electrons. The magnitude
$|U_{s}|$ is determined by competition between the electronic gain
of energy and its loss $\sim U_{s}^{2}$ from the lattice deformation
and charge redistribution (recall Sect. \ref{sec:gs} and Appendix
\ref{sec:e-i}). $3D$ ordering of signs $U_{s}=\pm |U_{s}|$
discriminates the FE and AFE states.

A conceptual conclusion is that the Mott-Hubbard state can be energetically favorable,
then the system will mobilize an available auxiliary parameter to reach it. In our case,
this parameter is the site disproportionation, which very fortunately is registered as
the ferroelectricity.

\section{Elementary excitations. \label{sec:excitations}}

\subsection{Solitons. \label{sec:solitons}}
\subsubsection{Solitons seen in most common cases. \label{sec:solitons-common}}
\begin{figure}[htb]
\centering\includegraphics*[width=0.7\textwidth]{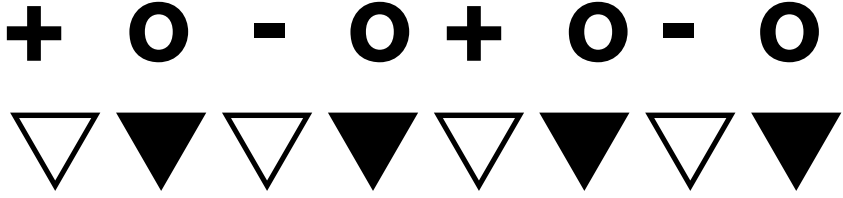}
\caption{Degeneracy scheme for electrons on a lattice with the site
dimerization. Lower row: white/black triangles show the good/bad
sites. Upper row: signs $\pm$ and zeros of the electronic wave
function at $E_F$. Shown configuration: $\varphi=0$ or $\pi$:
$\Psi=\pm 1$ or $\Psi=\mp 1$ at good sites and $\Psi =0$ at bad
sites. } \label{fig:signs}
\end{figure}

\begin{figure}[htb]
\centering
\includegraphics*[width=.4\textwidth]{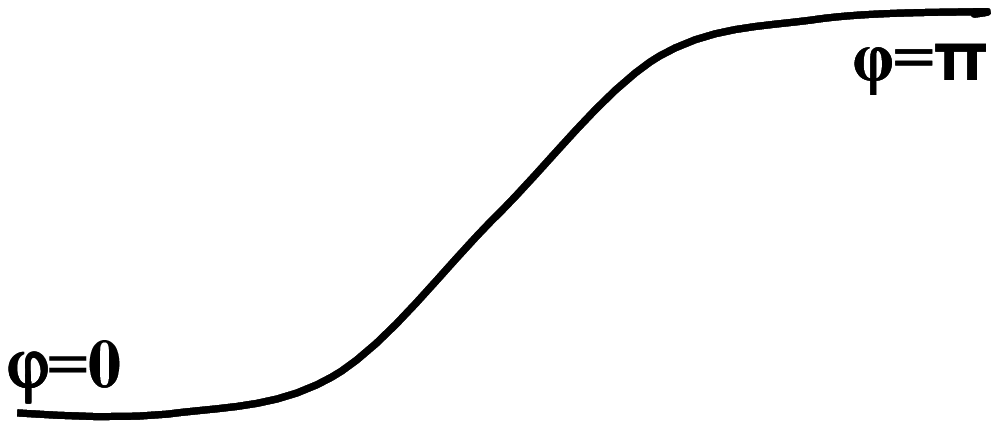}
\qquad
\includegraphics*[width=.4\textwidth]{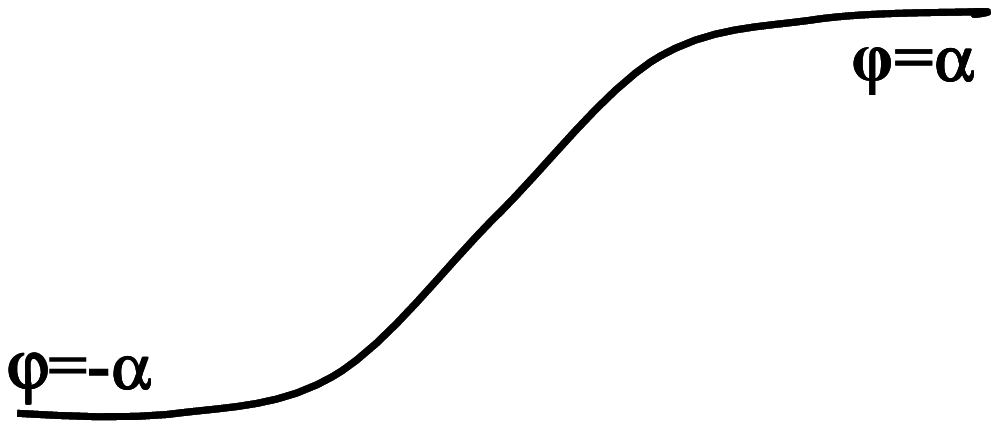}
\caption{Phase profiles $\varphi(x)$ of charged solitons. Left: $\pi$ solitons =
(anti)holons; spinless charge $\pm e$ particles seen in conductivity. Right: $\alpha$
solitons = domain walls of the ferroelectric polarization, seen in dispersion of the
permittivity.} \label{fig:solitons}
\end{figure}

For a given $U_{s}$, the ground state is doubly degenerate between
$\varphi=\alpha$ and $\varphi=\alpha \pm\pi $, which allows for
phase $\pm\pi$  solitons \cite{phi-particle} with the energy
$E_{\pi}=\Delta$; they are the charge $\pm e$  spinless particles -
the (anti)holons. Fig. \ref{fig:signs} illustrates the origins of
the potential $U_s\sin(2\varphi)$ by showing the sequence of zeros
and extrema $\pm 1$ of the electron wave function $\Psi$ taken at
$E_{F}$. The upper row corresponds to one of two correct ground
states, energy minima at $\varphi=0$, $\pi $. The degeneracy
"$\Psi=\pm 1$ at good sites" gives rise to $\pi$- solitons as kinks
between these two signs of the wave function, Fig.
\ref{fig:solitons} - left.

The $\pi$ solitons always determine the conductivity at $T<T_\rho$.
Thus in $\mathrm{(TMTCF)_2X}$ they are observed in conductivity at
both $T\gtrless T_{\mathrm{CO}}$; in compounds without the built-in
dimerization - at $T<T_{\mathrm{CO}}$. An expected characteristic
feature of solitonic conductivity is its strong reduction and
enhanced activation energy for the conductivity in the interchain
direction as it has been observed in $\mathrm{(TMTTF)_2X}$
\cite{auban:iscom-03}. Dynamics of solitons have been accessed in
recent tunneling experiments \cite{latyshev} on incommensurate CDWs.
There are also other cases of CDWs where conducting chains are
modulated by counterion columns. One is the very first known CDW
compound, the Pt chain KCP - in the stoichiometric version
$\mathrm{K_{1.75}[Pt(CN)_4]1.5H_2O}$. It was shown theoretically
\cite{kcp}, with an interpretation of available experimental data,
that the "weak two-fold commensurability" induced by columns of
counterions gives rise to the physics of phase $\pi$ solitons. The
mechanism of spinless conduction by the e-charged $\pi$-solitons
might be universal to all systems with the charge ordering,
irrespective to the ordering pattern. Actually the $\pi$- soliton is
a 1D vision of a vacancy or an addatom in the Wigner crystal, which
is the most universal view of the charge ordering. We shall return
to this topics in the Sect. \ref{sec:optics-solitons}.

\subsubsection{Ferroelectric solitons. \label{sec:solitons-fe}}
In the charge ordered  phase, $U_s\ne 0$ gives rise to the
ferroelectric ground state, if the same $\alpha $ is chosen for all
stacks. The state is the two-sublattice AFE if $\alpha $ signs
alternate as in the $\mathrm{(TMTTF)_{2}SCN}$, and more complex
patterns of $\alpha$ are possible as it has been found in new
compounds \cite{ravy:iscom-03}.

Spontaneous $U_{s}$ itself can change the sign between different
domains of ferroelectric displacements. Then the electronic system
must also adjust the mean value of $\varphi$ in the  ground state
from $\alpha$ to $-\alpha $ or to $\pi -\alpha$. Hence the
ferroelectric domain boundary $U_{s}\Leftrightarrow -U_{s}$ requires
for the phase $\alpha$-solitons, Fig. \ref{fig:solitons}, of the
increment $\delta \varphi=-2\alpha $ or $\pi -2\alpha$, whichever is
smaller; it will concentrate the non integer charge $q$:
$q/e=-2\alpha/\pi$ or $q/e=1-2\alpha/\pi$ per chain.

Above the 3D ordering transition $T>T_{\mathrm{CO}}$, the
$\alpha$-solitons can be seen as individual particles, charge
carriers. Such a possibility requires for the fluctuational $1D$
regime of growing charge ordering. It seems to be feasible in view
of a strong increase of $\varepsilon$ at $T>T_{\mathrm{CO}}$ even
for the AFE case of the $\mathrm{X=SCN}$ in Fig. \ref{fig:eps-log},
which signifies the growing single chain polarizability before the
3D interactions enter the game.

\begin{figure}
  \includegraphics[width=0.7\textwidth]{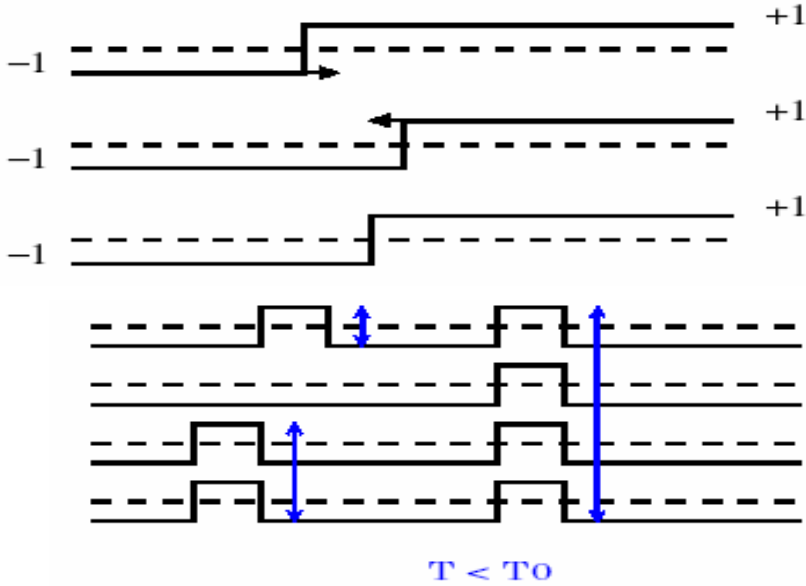}\\
\caption{Crossover from on-chain ferroelectric solitons into ferroelectric domain walls at
$T<T_{\mathrm{CO}}$. Upper three lines: aggregation of solitons into the wall. Lower three
lines: binding of solitons into pairs with subsequent aggregation into the bubble - the
nucleus of the opposite polarization. After \cite{teber}.}\label{fig:agregation}
\end{figure}

Below $T_{\mathrm{CO}}$, the $\alpha$- solitons must be aggregated
into domain walls \cite{teber}, separating domains of opposite
ferroelectric polarization. The noticeable asymmetry in the
frequency dependence of $\varepsilon^\prime$ above and below
T$_{\mathrm{CO}}$, Fig. \ref{fig:disp}, might be just due to this
aggregation. The nonsymmetry shows up much stronger in the frequency
dependence of the imaginary part of the permittivity
$\varepsilon^{\prime\prime}$, see figures in
\cite{nad-tau,nad:06,monceau:06}. The peak in
$\varepsilon^{\prime\prime}(\omega)$ determines the relaxation time
$\tau$ showing
two temperature regimes: \\
1) Near $T_{\mathrm{CO}}$, $\tau(T)$ grows sharply corresponding to the expected slowing
down of the critical collective mode; \\
2) At low $T$, $\tau$ grows exponentially following the activated law with the energy
similar to the one of the conductivity, $\Delta$; hence the relaxation is external, via
resistance of the electrically coupled gas of charge carriers - $\pi$- solitons.\\
3) Below the main peak in $\varepsilon^{\prime\prime}(\omega)$, a long tail appears at
$T<T_{\mathrm{CO}}$, showing a weak secondary maximum. This tail may be well interpreted
as the internal relaxation of the ferroelectric polarization through the motion of
pinned domain walls - aggregated $\alpha$- solitons.

\subsection{Effects of subsequent transitions:\\
 Spin-charge reconfinement and combined solitons. \label{sec:solitons-combined}}

Physics of solitons is particularly sensitive to a further symmetry
lowering, and the subsequent anion ordering of the tetramerization
in $\mathrm{(TMTTF)_{2}ReO_{4}}$ \cite{anions,nad:02,brazov:ICSM-02}
is a very fortunate example. This true present from the Nature
demonstrates clearly the effect of the spin-charge reconfinement.
Similar effects are expected for the Spin-Peierls state, e.g. in
$\mathrm{(TMTTF)_2PF_{6}}$, but the clarity of the 1D regime in
ReO$_{4}$ is uniquely suitable to keep the physics of solitons on
the scene.

The conductivity plot for the  ReO$_{4}$ case in Fig.
\ref{fig:ReO4+SCN} shows, at T=T$_{AO}$, the jump in $\Delta $
(actually even in G(T), see details in \cite{nad:02,nad:06}) which
is natural for the first order transition. We argue that the new
higher $\Delta $ at T$<$T$_{AO}$ comes from special topologically
coupled solitons which explore both the charge and the spin sectors.

Now we must invoke also the phase $\theta$ describing the spin $\sigma$ degree of
freedom, such that $\theta^\prime/\pi $ is the smooth spin density. Its free distortions
are described by the Hamiltonian
\begin{equation}
H_\theta=\frac{1}{4\pi}\left\{(\partial_{x}\theta)^{2}v_{\sigma}+
(\partial_{t}\theta)^{2}/v_{\sigma} \right\} \label{H-theta}
\end{equation}
where $v_{\sigma}\sim v_{F}$ is the spin sound velocity. (The absence of the factor
$\gamma^{-1}$ in \ref{H-theta} in comparison with (\ref{H-phi}) is not a typing
mistake.)

\begin{figure}[htb]
\centering\includegraphics*[width=0.7\textwidth]{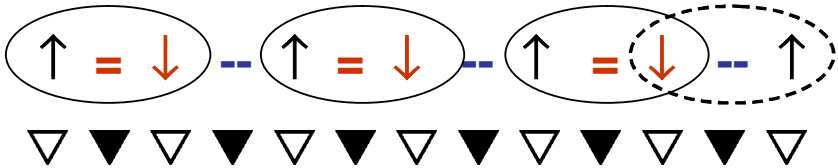} \caption{Effect of the
tetramerization leading to the energy (\ref{H-V}). White and black triangles show the site
inequivalence after the charge ordering. Tetramerization adds the inequivalence of
spin-exchange bonds {\bf =} and {\bf --}. Ellipses show the preferable spin singlets. The
last dashed ellipse shows an alternative spin configuration which was equivalent above the
$T_{AO}$; now, at $T<T_{AO}$, it becomes the spin excitation.}
 \label{fig:sites}
\end{figure}

The additional deformation of tetramerization, Fig. \ref{fig:sites}, acts upon electrons as a $2K_{F}$ CDW,
Fig. \ref{fig:umklaps}- right panel, thus adding the energy
\begin{equation}
H_V=-V\cos(\varphi-\beta)\cos\theta \label{H-V}
\end{equation}
(Here the shift $\beta$, mixing of site and bond distortions, reflects the lack of the
inversion symmetry, since $T_{AO}$ is already below $T_{\mathrm{CO}}$.) Within the
reduced symmetry of Fig. \ref{fig:sites}, the invariant Hamiltonian becomes

\begin{equation}\label{H-U+V}
H_U+H_V= -Ucos(2\varphi -2\alpha) - Vcos(\varphi -\beta)cos\theta
\end{equation}

Fig. \ref{fig:sites} suggests a schematic illustration for the
effect of the tetramerization. Inequivalence of bonds \textbf{=} and
\textbf{--} between good sites endorses ordering of spin singlets.
Also it prohibits translation by the two-site distance which was
explored by the $\delta \varphi =\pi $ soliton. But its combination
with the soliton ($\delta\theta=\pi$,  which shifts the sequence of
singlets) carrying the unpaired spin, is still allowed as the
selfmapping.

Formation at T$<$T$_{AO}$ of the new V-term (\ref{H-V}) destroys the
spin liquid, which existed at T$>$T$_{AO}$ on top of the charge
ordered state. The $V$ term in (\ref{H-U+V}) lifts the continuous
$\theta$- invariance, thus opening at $T<T_{AO}$ the spin gap
$\Delta_{\sigma}\sim V^{2/3}$, as it is known for the spin-Peierls
transitions \cite{spin-p,reidel-fukuyama}. Moreover, it lifts even
the discrete invariance $\varphi\rightarrow\varphi +\pi $ of
$H_{U}$, thus prohibiting the $\pi$ solitons to exist alone; now
they will be confined in pairs (either neutral or 2e - charged)
tightened by spin strings. But the joint invariance
\[
\varphi\rightarrow\varphi +\pi \ ,\ \theta\rightarrow\theta +\pi
\]
is still present in (\ref{H-V}), giving rise to \emph{compound topological solitons}
\cite{topdef} (cf. \cite{reidel-fukuyama}).

The major effects of the tetramerization V-term are the following:

a) to open the spin gap 2$\Delta_{\sigma}$ corresponding to creation
of new \{$\delta\theta =2\pi$, $\delta\varphi=0$\} uncharged
spin-triplet solitons,

b) to prohibit former $\delta\varphi =\pi $ charged solitons, the
holons, -- now they are confined in pairs bound by spin strings, the
activation $\gtrsim 2\Delta$ is required.

c) to allow for topologically bound spin-charge compound solitons
$\{\delta\varphi=\pi\}$, $\delta\theta=\pi$, which leave the
Hamiltonian (\ref{H-U+V}) invariant, the activation $\gtrsim\Delta$
is required.

\begin{figure}[htb]
\centering\includegraphics*[width=0.7\textwidth]{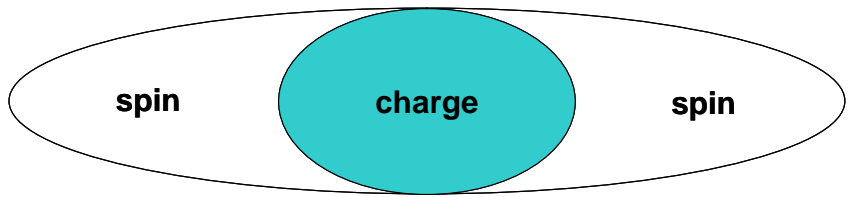}
\caption{Illustration of confinement below the tetramerization
transition. Different scales of spin and charge distributions within
the compound solitons: charge e, $\delta\varphi=\pi $ is
concentrated sharply within $\xi_{\rho}\sim\hbar v/\Delta$; spin
$1/2$, $\delta\theta=\pi$
 is concentrated loosely within $\xi_{\sigma}\sim\hbar v/\Delta_{\sigma}$.}
\label{fig:confin}
\end{figure}

For the last compound particle (c), the quantum numbers are as for a
normal electron: the charge $e$ and the spin $1/2$, but their
localization is very different, Fig. \ref{fig:confin}. The compound
soliton is composed with the charge $e$ core soliton (here $\delta
\varphi =\pi$ within the shorter length $\xi_{\rho}\sim \hbar
v_{F}/\Delta$) which is supplemented by spin $1/2$ tails of the spin
soliton (here $\delta\theta =\pi $ within the longer length
 $\xi_{\sigma}\sim\hbar v_{F}/\Delta_{\sigma}\gg \xi_{\rho}$).
  This complex of two topologically bound solitons gives the carriers seen
at $T<T_{AO}$ at the conductivity plot for the $\mathrm{X=ReO_4}$, Fig.
\ref{fig:ReO4+SCN} and \cite{brazov:ICSM-02}.

Similar effects should take place below intrinsically electronic transitions,
particularly relevant can be the spin-Peierls one for $\mathrm{X=PF_6}$. But there the
physics of solitons will be shadowed by $3D$ electronic correlations, which are not
present yet for the high $T_{AO}$ of $\mathrm{X=ReO_4}$ case.

\section{Optics \label{sec:optics}}
\subsection{Optics: collective and mixed modes.  \label{sec:optics-coll}}

\subsubsection{Optics: phase mode.}
Beyond the ferroelectric phase (e.g. the AFE state, or the
disordered one at $T>T_{\mathrm{CO}}$) the optical absorbtion starts
at the frequency $\omega_{t}<2\Delta$, which is the bottom of the
spectrum of phase excitations
$\omega^{2}=(v_{\rho}k)^{2}+\omega_{t}^{2}$. In the quasi-classical
limit of small $\gamma\ll 1$, $\omega_t$ can be interpreted as the
frequency of phase oscillations of the $\mathrm{4K_F}$ CDW around
the minimum of the commensurability energy (\ref{H-U}). The
renormalized value $U^{\ast}$ of $U$ in (\ref{H-U}) can be expressed
through this observable parameter as

\begin{equation}\label{U-omega_t}
U^{\ast}=\frac{\omega_{t}^{2}}{8\pi\gamma v_{\rho}} \,,\;
 \omega_{t}\approx \pi\gamma\Delta<\Delta  \;\; (\gamma\ll 1)
\end{equation}
As a ''transverse'' frequency of the optical response, $\omega_{t}$ gives the background
dielectric susceptibility:
\begin{equation}
\varepsilon_{\Delta}(\omega) =\frac{\omega_{p}^{*2}}{(\omega_{t}^{2}-\omega^2)} \,,\;
\omega_{p}^{*2}=\gamma\frac{v_{\rho}}{v_{F}}\omega_{p}^{2}  \,,\;
\omega_{p}^{2}=8\frac{e^{2}v_{F}}{\hbar s} \label{eps-omega_t}
\end{equation}
where $\omega_{p}^{*}$ and $\omega_{p}$ are the actual plasma
frequency of the parent metal and its bare value, $s$ is the area
per stack.  In exact theory of the quantum sine-Gordon model,
$\omega_{t}$ appears as the fist bound state $E_1$ of the pair of
solitons, see \cite{klainert:79,reidel-maki,essler:01-pr}. Its exact
relation to the gap $2\Delta$ is
$\omega_{t}=E_1=2\Delta\sin(\pi\tilde\gamma/2)$ as explained below
in Eq.(\ref{E_n}).

\subsubsection{Optical permittivity $\varepsilon(\omega)$ near the ferroelectric transition.}

\begin{figure}
  \centering
  \includegraphics*[width=0.7\textwidth]{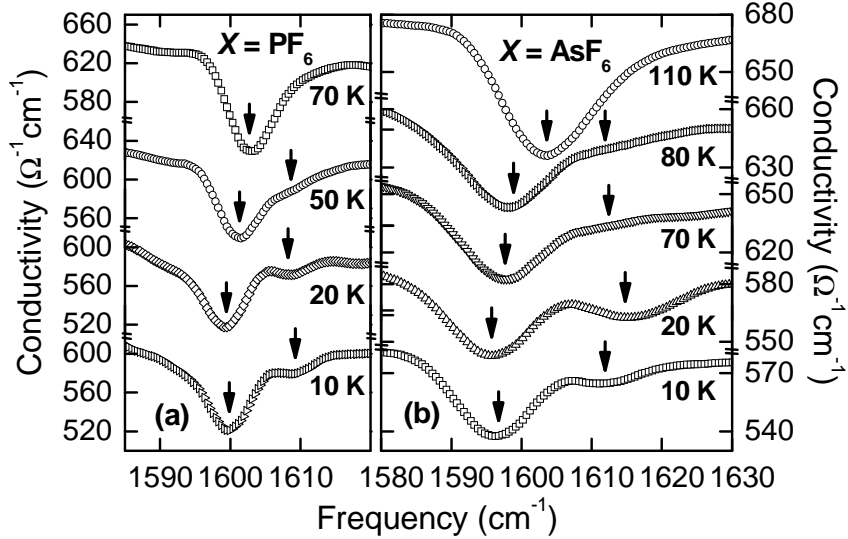}\\
  \caption{CO effect upon the molecular mode. The figure demonstrates two effects:
   1) Fano antiresonance at any $T\gtrless T_{\mathrm{CO}}$ which might be due the stack
   polarizability;
   2) its splitting at $T<T_{\mathrm{CO}}$ due the CD - in accordance with the NMR.
   After \cite{dumm:ecrys-05}.}
   \label{fig:dumm}
\end{figure}

We shall present, without derivation, the formula for the mixed electron-phonon
contribution to the permittivity, valid at $T\geq T_{\mathrm{CO}}$ - above and
approaching the ferroelectric transition:
\begin{equation}
\frac{\varepsilon (\omega)}{\varepsilon_{\infty}}= 1+\frac{(\omega_{p}^{\ast}/\omega
_{t})^{2}(1-(\omega /\omega_{0})^{2})}{(1-(\omega /\omega_{0})^{2})(1-(\omega /\omega
_{t})^{2})-Z}\;\;,\ Z=\left(\frac{\omega_{cr}}{\omega_{t}}\right)^{2-4\gamma}\leq 1
 \label{epsilon}
 \end{equation}
Here $\omega_0$ is a frequency of the molecular mode coupled to the charge ordering;
$\omega_{cr}(T)$ is the critical value of the collective mode frequency $\omega_{t}(T)$,
below which the spontaneous ferroelectric charge ordering  takes place.
 Near the critical point $Z(T_{\mathrm{CO}})=1$ we see here:\\
\begin{tabbing}
combined electron-phonon resonance \= Hierarchy of phases in quasi 1D organic
conductors. \kill
\emph{Fano antiresonance} \>\ at $\omega_{0}$, just as observed in \cite{dumm:ecrys-05}
 - Fig. \ref{fig:dumm}\\
\emph{Combined electron-phonon resonance} \>\ at
$\omega_{0t}^{2}\approx\omega_{0}^{2}+\omega_{t}^{2}$\\
\emph{FE soft mode} \>\ at
$\omega_{fe}^{2}\approx(1-Z)(\omega_{0}^{-2}+\omega_{t}^{-2})^{-1}$
\end{tabbing}

Near $T_{\mathrm{CO}}$ the ferroelectric mode might be overdamped; then it grows in
frequency following the order parameter, that is as $\omega_{fe}\sim \varepsilon^{-1/2}$
(which can yield two orders of magnitude at low $T$) to become finally comparable with
$\min{\{\omega_{0},\omega_t\}}$. Suggesting smooth $T$ dependencies $\omega_{t}(T)$ and
$\omega_{cr}(T)$, we find the critical singularity at $\omega =0$ as $\varepsilon
(T)=A|T/T_{\mathrm{CO}}-1|^{-1}$. It develops upon the already big gapful contribution
$A\sim(\omega_{p}^{\ast}/\omega_{t})^{2}\sim 10^{3}$ in agreement with experimental
values $\varepsilon\sim10^{4}T_{\mathrm{CO}}/(T-T_{\mathrm{CO}})$. All that confirms
that the ferroelectric polarization comes mainly from the electronic system, even if the
corresponding displacements of ions are very important to choose and stabilize the long
range $3D$ order.

\subsection{Optics: solitons.  \label{sec:optics-solitons}}
Thanks to detailed information on the sine-Gordon model, we can clearly formulate the
expectations for optical properties related to physics of solitons (see details in
\cite{klainert:79}, and \cite{reidel-maki} for a review; the contemporary stage and
comparison of different approaches can be found in \cite{essler:01-pr}).

The most general feature of the optical spectra, valid through the
whole gapful regime $\gamma<1$, is the two-particle gap
$E_{g}=2\Delta$ corresponding to the creation of a pair of $\pm\pi$
solitons. Contrary to the common sense intuition and the elementary
theory, the absorption $I(\omega)$ has no singularity at the
threshold $E_g$. The optical density of states law
$DOS\sim(\omega-E_g)^{-1/2}$ is compensated by vanishing of the
matrix element. As a result, the absorption starts smoothly as
$I\sim(\omega-E_g)^{1/2}$. Realistically, the gap will be seen only
as the nonsingular threshold for the photoconductivity, where other
absorption mechanisms are excluded. For the built-in Mott state
without the charge ordering, $1/2<\gamma<1$, there is no absorption
below E$_{g}$. This case was studied in detail in
\cite{essler:01-prl}.

But for the Mott state due to the spontaneous charge ordering,
$\gamma<1/2$, there might be also sharp peaks of the optical
absorption even below the two-particle gap $E_{g}=2\Delta$. Actually
the spectral region $\omega<2\Delta$ is filled by a sequence of
quantum breathers, bound states of two solitons at energies
\begin{equation}
E_n=2\Delta\sin(\frac{\pi}{2}\tilde\gamma n)\; , \;
\tilde\gamma^{-1}=\gamma^{-1}-1. \label{E_n}
\end{equation}
Here the substitution of $\gamma$ for $\tilde\gamma$ takes into
account quantum corrections to the control parameter itself
\cite{reidel-maki}. The modes with odd numbers $n$ are optically
active. The primary, lowest bound state $E_1$ gives the corrected
value for the collective mode $\omega_t$. It reduces to the
classical result of (\ref{U-omega_t}) in the limit of small
$\gamma\ll 1$, where
$E_1=2\Delta\sin(\pi/2\tilde\gamma)\approx\pi\gamma\Delta=\omega_{t}<2\Delta$

At a sequence of special values of $\gamma_n=1/2n$, starting from
$\gamma=1/2$ downwards, the next bound state splits off from the
gap. Only at these moments, the absorption edge singularity blows up
from the smooth dependence $\sim(\omega-E_g)^{1/2}$ to the
divergency $DOS\sim(\omega-E_g)^{-1/2}$. (There is a crossover near
these values of $\gamma$ \cite{essler:01-pr}, where a rounded edge
singularity can be observed.) This property opens an amusing
possibility to observe spectral anomalies at special values of the
monitoring parameter, varying it e.g. by applying pressure.

We see that the scheme changes qualitatively  just at the borderline
for the charge ordering  instability $\gamma=1/2$: from the
essentially quantum regime $1/2<\gamma<1$ (with $E_{g}=2\Delta$ and
no separate $\omega_t$ at all) to the quasi classical low $\gamma$
regime with a sequence of peaks between $2\Delta$ and $\omega_t$.
This fact was not quite recognized in existing interpretations, see
\cite{giamarchi:iscom-03}, of intriguing optical data
\cite{vescoli:98,vescoli:99,schwartz:98}.  E.g. the detailed
theoretical work \cite{essler:01-prl} was all performed for the case
equivalent to $\gamma>1/2$, in our notations, and cannot be applied
to the $\mathrm{(TMTTF)_{2}X}$ case as it was supposed. The studies
of \cite{klainert:79} and \cite{essler:01-pr} are quite applicable,
with the adjustment to the two-fold commensurability, which we have
implied above.

\begin{figure}[htb]
\centering
\includegraphics*[width=0.7\textwidth]{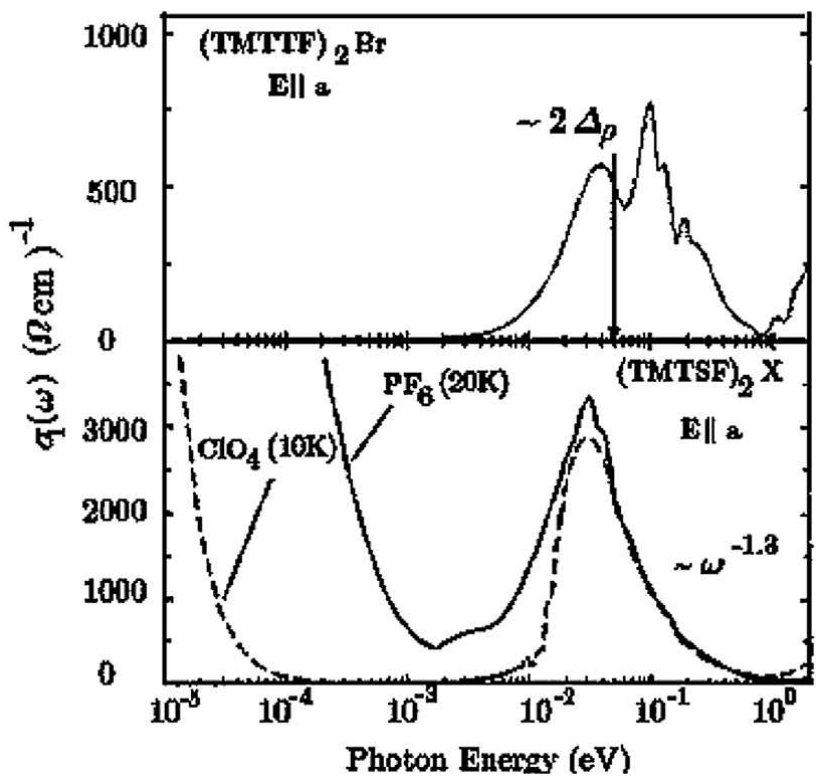}
\caption{Comparison of optical absorption of TMTTF and TMTSF
compounds in a wide $\omega$ range. It seems that after subtracting
molecular peaks at the high $\omega>2\Delta_\rho$ slope (upper
panel) the gaps shapes will be quite similar in both TMTTF and TMTSF
cases. Is it a way to prove the hidden charge ordering  in the
metallic state of the Se subfamily? After \cite{vescoli:98}}
\label{fig:optics-comp}
\end{figure}

\begin{figure}[htb]
\centering\includegraphics*[width=0.7\textwidth,height=10cm]{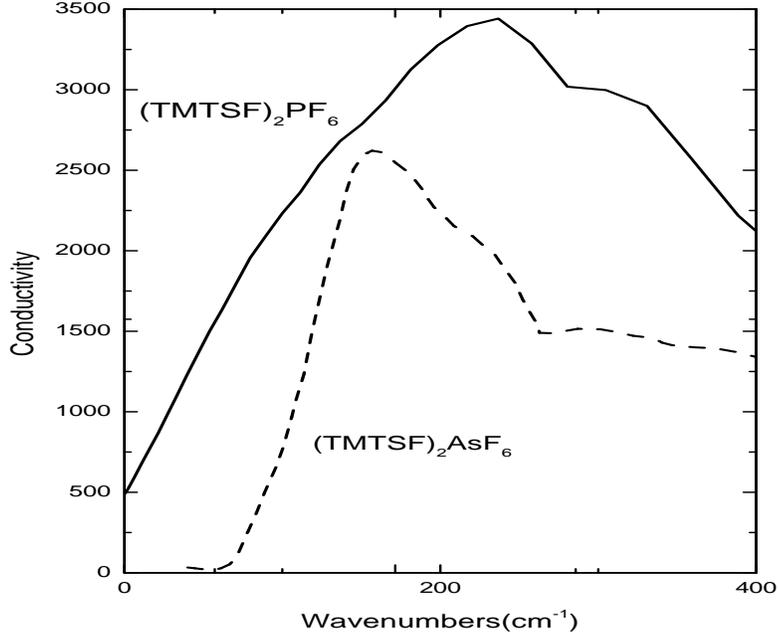}
\caption{Optical spectra of two $\mathrm{(TMTSF)_2X}$ salts. The
case of AsF6 shows a clear excitonic peak which may be the mode
$\omega_t$. The case of $\mathrm{X=PF_6}$ shows a more smeared
feature, probably due to the noticeably high SDW gap within the
charge ordering  one. The shoulders seen at frequencies above the
peaks can be well interpreted as the two-particle gaps $2\Delta$.
Then they will determine the thresholds in photoconductivity. M.
Dressel, unpublished.} \label{fig:pf6+asf6}
\end{figure}

The available experiments, see Figs. \ref{fig:optics-comp} and
\ref{fig:pf6+asf6}, can be interpreted as observations of the
collective mode - solitons' bound state. Photoconductivity
experiments (distinguished from the bolometric effect !) are
necessary to discriminate between two absorption mechanisms: at
$\omega_t$ and above $2\Delta$.

Recall that quenching of the edge singularity is not just a peculiar
property of the sin-Gordon model. It takes place also in theory
\cite{kirova:02} of generic 1D semiconductors such as conjugated
polymers, as a consequence of the final state interaction via the
long range Coulomb forces. Merging of local and Coulomb interaction
is still an unattended issue in theory of solitons.

\subsection{Optics: summary.  \label{sec:optics-summ}}

Now we summarize shortly the main expected optical features \cite{brazov:ecrys-02}.

\textbf{I.} \underline {\textit{In any case of the CO}}, for both FE or AFE orders, we
expect

Ia) Strongest absorption feature comes from the phase mode, an

analogy of the exciton as the bound kink-antikink pair at $\omega_t$;

Ib) Two-particle gap 2$\Delta$ (e.g. the photoconductivity) lies
higher than $\omega_t$,

it is given by free pairs of $\pi$-solitons, $\Delta=E_\pi$;

Ic) Spectral region $\omega_{t}<\omega<$2$\Delta $ may support also
quantum breathers --

higher bound states of solitons.

\textbf{II.} \underline {\textit{In case of the ferroelectric order}} we expect
additionally or instead of I:

IIa) Fano antiresonance at the bare phonon mode frequency $\omega_0$

coupled to the CO;

IIb) Combined electron-phonon resonance at $\omega_{0t}>\omega_{0},\omega_{t}$,

which substitutes for Ia;

IIc) Ferroelectric soft mode at $\omega_{fe}$, evolving from $\omega_{fe}=0$  at
T$<$T$_{\mathrm{CO}}$.

Unfortunately, the optical picture of $\mathrm{(TMTTF)_2X}$ compounds is complicated by
multiple phonon lines filling just the relevant region of the optical gap. But this
complication may not be all in vain, if it is viewed also as another indication for the
charge ordering. Indeed, surprisingly (kept noticed since early 80's \cite{phonons})  high
intensity of molecular vibrations in TMTTF  may be just due to the inversion symmetry
lifting by the charge ordering, recall \cite{rice-phonons} for well confirmed mechanisms of
phonons activation by the CDW formation. Oppositely, weaker phonon lines in TMTSF
\cite{phonons} may speak in favor of a fluctuational regime of the charge ordering  in
accordance with the pseudogap, rather than a true gap, in electronic optical transitions.

The whole obstacle can be overcome by experiments on low gap charge ordered  states like
in $\mathrm{(TMTTF)_{2}Br}$, or under pressure. There, as one can guess from other
experiments \cite{nagasawa:ISCOM-03,pressure}, the gap value will fall below the region
of intensive molecular vibrations, which today prevent the observations. Fig.
\ref{fig:optics-comp} shows that it is happening indeed, and just for X=Br as expected:
the electronic spectrum starts to split off from the vibrational one.

Recall finally the great experience of optics in another type of
strongly correlated 1D electronic systems: conjugated polymers. They
were studied by such a complex of optical techniques as
photoconductivity, stimulated photoemission, photoinduced
absorption, electro-absorption, time resolved measurements, see
refs. in \cite{kirova:04}.

A popular interpretation (see \cite{giamarchi:iscom-03} for a
review) of optics \cite{vescoli:98,vescoli:99,schwartz:98} and
sometimes conductivity \cite{coulon:iscom-03} in TMTCF - type
compounds neglects the dimerization (either generic from bonds or
from sites via the charge ordering) and relies upon the weaker 4-
fold commensurability effects. They give rise to the energy $\sim
U_{4}\cos 4\varphi$ originated by higher order (8 particles
collisions) Umklapp processes; its stabilization would require a
stronger $e-e$ repulsion (or slowing down, see Sect.
\ref{sec:generic}) corresponding to $\gamma <1/4$. While not
excluded in principle, this mechanism does not work in TMTTF case,
already because this scenario does not invoke any charge ordering
instability. Moreover, the experiment shows, Fig. \ref{fig:cond},
that even small increments of the dimerization, just below the
second order transition at $T_{\mathrm{CO}}$, immediately transfer
to the activation energy, hence the domination of the two-fold
commensurability.

\section{Fate of the metallic TMTSF subfamily.  \label{sec:tmtsf}}

By now, the re-valuation  definitely concerns mostly the TMTTF
subfamily, whose members usually, by the temperature
$T_{\mathrm{CO}}$, already fall into the charge-gap regime starting
at $T_\rho>T_{\mathrm{CO}}$, see Fig. \ref{fig:cond}. The TMTSF
compounds are highly conductive, which does not allow for
measurements of the low frequency permittivity $\varepsilon$. Also
the  NMR splitting \cite{brown,fujiyama:02-03} will probably be
either too small or broadened because of expected disordered or
temporal nature of the charge ordering  in the metallic phase (see
\cite{takahashi:ecrys-05} for a proved example of another system
with the charge ordering). Nevertheless the transition may be there,
just being hidden or existing in a fluctuation regime, as may be
realized, e.g., in stripe phases of High-$T_{c}$ cuprates
\cite{stripes}. If and when confirmed, the entire picture of
intriguing abnormal metallic state \cite{jerome:99} will have to be
revised following the TMTTF case.

The signature of the ferroelectric charge ordered  state may have
been already seen in optical experiments
\cite{vescoli:98,vescoli:99,schwartz:98}, as we have discussed
above. Indeed, the Drude-like low frequency peak appearing within
the pseudogap, Fig. \ref{fig:optics-comp} - lower panel, can be
interpreted now as the optically active mode of the ferroelectric
polarization; the joint lattice mass, see Sect. \ref{sec:generic},
will naturally explain its otherwise surprisingly low weight. Recall
here the earlier conjectures on collective nature of the
conductivity peak \cite{jerome:94,jerome:04,timusk}, derived from
incompatibilities of IR optics, conductivity and  NMR. Vice versa,
the ferroelectric mode must exist in TMTTF compounds, whose
identification is the ultimate goal. Even the optical pseudogap
itself \cite{vescoli:98,vescoli:99,schwartz:98}, being unexpectedly
large for TMTSF compounds, Fig. \ref{fig:optics-comp}-top, with
their less pronounced dimerization of bonds, can be largely due to
the hidden spontaneous dimerization of sites. Even the shapes of the
gap and the pseudogap in TMTTF and TMTSF cases appear to look
similar, if the first one is cleaned from molecular modes, Fig.
\ref{fig:optics-comp} - upper panel. Optical experiments will
probably be elucidated when addressed to members of the
$\mathrm{(TMTTF)_2X}$ family, showing the charge ordering  with a
particularly reduced value of the associated gap (below typical
molecular vibrations - down to the scale of the pseudogap in
$\mathrm{(TMTSF)_2X}$).

Hidden existence of the charge ordering  and the local ferroelectricity, at least in
fluctuating regime, is the fate of the TMTSF compounds and the major challenge is to
detect it, as it was seen explicitly in TMTTF compounds.  Key effects of anion
orderings, particularly the opportunity to compare relaxed and quenched phases of
$\mathrm{(TMTSF)_2X}$, also are waiting for attention.

\section{Origin and range of basic parameters. \label{sec:theory2}}
\subsection{Generic origins of basic parameters:
\newline
Interactions among electrons or with phonons? \label{sec:generic}}

The combination of optical and conductivity data can provide a
deeper insight into the nature of observed regimes. The value of
$\Delta$ is already known as the conductivity activation and
$2\Delta$ can be found independently as the photoconductivity
threshold; $\omega_{t}$ is measurable through the optical
absorption. Then their ratio will provide the basic microscopic
parameter $\gamma$.

The full quantitative implementation requires to resolve for divergence (2,3 or even more
times !) \cite{schwartz:98,phonons} in values of such a basic and usually robust parameter
as the plasma frequency, if it is extracted from different parts of the spectrum. This
discrepancy can signify the strong renormalization $\omega_{p}^{\ast}\ll\omega_{p}$, which
can develop while the probe frequency decreases from the bare scale $\omega_{p}>1eV$ to the
scale $\omega_{t},\omega_{0}\sim 10^{-2}eV$.

Remind the full (kinetic $\sim C_{kin}$ and potential $\sim C_{pot}$) energy of elastic
deformations for the charge phase $\varphi$:

\[
\frac{\hbar v_{F}}{4\pi}
 \left\{(\partial_{t}\varphi)^{2}C_{kin}/v_{F}^{2}+(\partial_{x}\varphi)^{2}C_{pot}\right\}
\equiv \frac{1}{4\pi\gamma}\left\{(\partial_{t}\varphi)^{2}/v_{\rho}
+(\partial_{x}\varphi)^{2}v_{\rho}\right\}
\]
where
\[
\gamma =\frac{1}{C_{pot}C_{kin}}\, , \, \frac{v_{\rho}}{v_{F}}=\frac{C_{pot}}{C_{kin}}\,
,\, \frac{\omega_{p}^{\ast}}{\omega_{p}}
=\left(\gamma\frac{v_{\rho}}{v_{F}}\right)^{1/2}=\frac{1}{C_{kin}}
\]

The potential parameter $C_{pot}$ is "1+ e-e repulsion contribution". The kinetic
parameter $C_{kin}$ is "1+lattice adjoined mass". The parameter $\gamma$ contains a
product of C's -- not distinguishable separately; the velocity $v_\rho$ contains a ratio
of C's  -- not distinguishable separately. But $\omega_p^*$ contains only $C_{kin}$,
which gives a direct access to the joint lattice dynamics. (Another factor for reduction
of the parameter $\gamma$, the Coulomb hardening $C_{pot}>1$, acts upon $\gamma$ and
velocity $v_{\rho}$\footnote{In CDWs the Coulomb hardening, as well as the very common
mass enhancement \cite{mass}, are confirmed experimentally \cite{harden}}, but cancels
in their product which gives $\omega_{p}^{\ast}$.) The lowering of $\omega_{p}^{\ast}$
singles out the effect of the effective mass enhancement $C_{kin}>1$, which is due to
coupling of the charge ordering  with $4K_F$ phonons \cite{fin-braz}\footnote
 {X-ray scattering gives a direct evidence for the coupling of the $4K_F$ electronic density
 with either lattice displacements \cite{pouget-4kf,kagoshima,4kf}
 or intramolecular modes \cite{nogami:99}.}.

The final step is to notice that the mass enhancement will not be
effective above the $4K_F$ phonon frequency $\omega_{ph}$; actually
$C_{kin}$ is a function of $\omega$:
$C_{kin}(\omega)=C_{kin}(0)\omega_{ph}^2/(\omega_{ph}^2+\omega^2)$.
It explains the difference in values of $\omega_{p}^{\ast}$
extracted from high and from intermediate frequency ranges.  If
true, then the charge ordered  state is a kind of a polaronic
lattice (as it was already guessed for some CDWs
\cite{polaronic-CDW}). It resembles another Wigner crystal:
electrons on the liquid Helium surface, see \cite{shikin}, where
selftrapped electrons gain the effective mass from surface
deformations - the ripplons.

\subsection{Where are we?  \label{sec:where?}}
There are still several fascinating questions to understand:
 \newline Why the charge ordering  is so common and appears at such a high
energy/temperature scale?
\newline Why do we see it instead of the abnormal metal
(Luttinger Liquid regime)?
\newline Why does it develop spontaneously before the 1/4 filling
effects have a chance to be seen?

We may get an idea for the answers to these questions by analyzing
the sequence of various regimes, as it appears by changing the
control parameter $\gamma$.

Phenomenological Hamiltonian may have the following typical
components (we restrict it here to terms containing only the charge
phase $\varphi$).

$H_\varphi\sim\gamma^{-1}[v_{\rho}(\partial_{_{^{x}}}\varphi)^{2} + v_{\rho}^{-1}
(\partial_{t}\varphi)^{2}$] -

\qquad from the electron liquid or fluctuating $4K_F$ CDW (\texttt{basic})

+$U_{1}\cos(\varphi)$ \ -- from tetramerization or spin-Peierls (\texttt{spontaneous,
frequent})

$+U_{2}\cos(2\varphi)$ -- from dimerization  (\texttt{built-in or spontaneous, typical})

$+U_{3}\cos(3\varphi)$ -- from trimerization: TTF-TCNQ under
pressure,

\hfill  NMP-TCNQ (\texttt{special})\ \ \ \ \ \ \ \ \ \ \ \

$+U_{4}\cos(4\varphi)$ -- built-in, from host lattice (\texttt{typical}).

The parameter $\gamma=K_{\rho}$ controls quantum fluctuations of the phase $\varphi$
which will, or will not, destroy the above nonlinearities; it defines:

1. survival, against renormalization to zero, of nonlinearities $\sim U_i$;

2. their spontaneous generation -- known for $U_1,U_2$;

3. physics of solitons and the collective mode;

4. relevance of the interchain coupling: metal/insulator branching.

We can quote the following regimes.

$\gamma<1$ : Renormalized $U_2\neq 0$; the charge gap is originated
in case of built-in dimerization (of either bond or site types).
This is the generic Mott-Hubbard state, any repulsion is sufficient
to stabilize it. Solitons = holons appear as free excitations,
giving both the thermal activation energy in conductivity, and the
optical absorption threshold. This regime is certainly valid for all
$\mathrm{(TMTTF)_2X}$. It is not applicable to nondimerized
materials like DMtTTF and EDT-TTF-CONMe salts
\cite{batail,heuze:03}. They still would be metallic, unless falling
into the next regime described below, which does happen actually.

$\gamma<1/2=0.5$ :  $4K_F$ anomaly appears in X-ray scattering at high T, as registered
in fractionally filled cases  of TTF-TCNQ and its derivatives
\cite{4kf,kagoshima,NMP-TCNQ}. In 1/4 filled compounds the spontaneous site dimerization
potential $\sim U_s$ is formed, hence no need for a bare commensurability/umklapp
potential $\sim U_b$. This regime is proved to be valid, by observations of the
ferroelectric response and the NMR splitting, in most of
$\mathrm{TMTTF_2X}$, and by X-rays \cite{ravy:iscom-03} in $\mathrm{(DMtTTF)_2ClO_4}$.%
\footnote{The last compound, as well as $\mathrm{(EDT-TTF-CONMe)_2X}$, is still waiting for
the standard (ferroelectricity, NMR) characterization.}
  All materials are waiting for determining the optical signatures,
see Sect. \ref{sec:optics-summ}.

$\gamma<4/9=0.39$ : Trimerization lock-in of $4K_F$ + $2K_F$ superlattices (confirmed in
TMTTF-TCNQ under a special pressure \cite{jerome:78} and in NMP-TCNQ \cite{NMP-TCNQ}).

$\gamma<1/4=0.25$ : effects of $1/4$ filling may come to play (cf. the lock-in of
sliding of CDWs under stress in MX$_{3}$ conductors \cite{TaS3-stress,NbSe3-stress}).

$\gamma<\sqrt{5}-2\approx0.24$ : ultimate SDW instability, even for incommensurate
cases. (The confinement index of the electron pair in the course of the interchain
hopping is $(1/\gamma-\gamma)/2<2$ \cite{BY:phys-let,BY:jetp}). Seemingly, it is not the
case of TMTSF metals: they need the HMF support to form the FISDW.

$\gamma\approx0.23$ : a  guess from  high $\omega\gg\Delta$ tails
($\sim\omega^{-1/3}$, see Fig. \ref{fig:optics-comp}) of optical
absorption \cite{vescoli:99}, see also \cite{giamarchi:iscom-03}.

$\gamma<2/9\approx 0.22$ :  spontaneous trimerization which is not observed: TTF-TCNQ
needs a precise pressure to pinpoint exactly 1/3.

$\gamma<3-\sqrt{8}\approx0.17$ : last feature of electrons' Fermi surface disappears.
(The electron Green function index is $(\gamma+1/\gamma+2)/4<2$.) This regime was
guessed from ARPES experiments in $\mathrm{(TMTSF)_2X}$ \cite{arpes-tmtsf}, but not seen
in seemingly more correlated TTF-TCNQ \cite{arpes-tcnq}.

$\gamma=0.125$ : spontaneous $1/4$ filling in totally incommensurate chains. But the
usual CO, the $4K_F$ condensation, would has happened already well before.

\emph{Resume:} most of qualitative effects of electronic correlations would appear at
$\gamma<1/2$, where the system is already unstable with respect to the charge ordering. The
existing experiments on most studied materials pose the following constraints.
$\mathrm{(TMTTF)_2X}$ : $\gamma<0.5$; $\mathrm{(TMTSF)_2X}$ : $\gamma>0.24$; TTF-TCNQ :
$0.22<\gamma<0.39$.

\section{Conclusions and perspectives.  \label{sec:conclusions}}

We have presented the  key issues of related phenomena of the
Ferroelectricity and the Charge Ordering  in organic metals. In
$\mathrm{(TMTTF)_{2}X}$ the dielectric permittivity $\varepsilon$
demonstrates clear cases of the ferroelectric and anti-ferroelectric
phase transitions. The combination of conductivity and magnetic
susceptibility proves the spinless nature of charge carriers.  The
independence and occasional coexistence of structureless
ferroelectric transitions and usual anionic ones brings the support
of structural information. The sequence of symmetry breakings gives
access to physics of three types of solitons: $\pi$- solitons
(holons) are observed via the activation energy $\Delta $ in
conductivity; noninteger $\alpha$- solitons (ferroelectric domain
walls) provide the low frequency dispersion; topologically coupled
compound spin-charge solitons determine the conductivity below a
subsequent structural transition of the tetramerization. The
photoconductivity gap $2\Delta $ will be given by creations of
soliton - antisoliton pairs. The lower optical absorption comes from
the collective electronic mode: in the ferroelectric case it becomes
the mixed electron-phonon resonance coexisting with the Fano
antiresonance. The ferroelectric soft mode evolves from the
overdamped response at T$_{\mathrm{CO}}$. The reduced plasma
frequency signifies the slowing down of electrons' collective motion
by the adjoint lattice mass; it recalls that the charge ordering has
a $4K_F$ lattice counterpart in accordance with the X-ray experience
\cite{pouget-4kf}.

We propose that a latent charge ordering in the form of
ferroelectricity exists even in the Se subfamily
$\mathrm{(TMTSF)_{2}X}$, giving rise to the unexplained yet low
frequency optical peak and the enhanced pseudogap. Another,
interchain type of the charge disproportionation, known in the
relaxed $\mathrm{(TMTSF)_{2}ClO_{4}}$, is still waiting for
attention; possibly it is present, perhaps also in a hidden form, in
other superconducting cases. Discoveries of the ferroelectric
anomaly and of the related charge ordering call for a revaluation of
the phase diagram of the $\mathrm{(TMTTF)_{2}X}$ and similar
compounds and return the attention to the interplay of electronic
and structural properties. In this respect, we point to the old
theory \cite{BY:phys-let,BY:jetp} for the synergetic phase diagram
of these materials.

The ferroelectricity discovered in organic conductors, beyond its
own virtues, is the high precision tool to diagnose the onset of the
charge ordering and the development of its order. The wide range of
the ferroelectric anomaly (T$_{\mathrm{CO}}\pm $30K) suggests that
the growing charge ordering  dominates the whole region below and
even above these already high temperatures. Even higher is the
on-chain energy scale, from 500K up to 2000K, as given by the
conduction gaps formed at lower T, and by optical features. Recall
also the TTF-TCNQ with ever present, up to the RT, 4K$_{F}$
fluctuations. All that appears at the "\emph{Grand Unification}"
scale, which knows no differences with respect to interchain
couplings, anion orderings, ferro- and antiferroelectric types,
between Sulphur and Selenium subfamilies, between old faithful
incommensurate or weakly dimerized compounds and the new
quarter-filled ones. Hence the formation of the Electronic Crystal
(however we call it: \textit{disproportionation, ordering,
localization or Wigner crystallization of charges; } $4K_{F}$
\textit{density wave, etc}.) must be the starting point to consider
lower phases and the frame for their properties.

On the theory side, the richness of symmetry-defined effects of the
charge ordering, ferroelectricity, AFE and various AOs (see
\cite{BY:phys-let,BY:jetp} for earlier stages) allows for efficient
qualitative assignments and interpretations.

Still there are standing questions on charge ordering  and ferroelectricity:

Why is the charge ordering so common? \hskip1cm  Is it universal?

Why is the astonishing ferroelectricity so frequently encountered

\hfill as a form of charge ordering? \ \ \ \ \ \ \ \

Why do the AFE and more complex patterns appear on other occasions?

Is it a spontaneously created Mott-Hubbard state?

Is it a Wigner crystal, if yes then of what: electrons or polarons?

Is it a $4K_F$ CDW: driven by electrons and stabilized by the lattice?

Role of anions: is there a key to the FE/AFE choice?

Are there other challenges ignored since decades? Examples: the plasma frequency mystery,
the special anion ordering structure for superconducting phase.

The last remark reminds us that the story of hidden surprises may
not be over. Another, interchain, type of the charge
disproportionation, known in the relaxed
$\mathrm{(TMTSF)_{2}ClO_{4}}$ (see more in \cite{anions}), is still
waiting for attention, possibly being present in a latent form in
other superconducting cases. This additional anion ordering can
contribute to the controversial physics of inhomogeneous state at
the SC/SDW boundary under pressure (see Chaikin, J\'erome, Heritier,
et al in this volume). The coexistence of charge ordering  and anion
ordering transitions in $\mathrm{(TMTTF)_{2}ReO_{4}}$ warns for this
possibility in other cases as well. If it is not found to be the
case, then we will be forced to accept two different types of
superconductivity in $\mathrm{(TMTSF)_{2}ClO_{4}}$ and in
$\mathrm{(TMTSF)_{2}PF_{4}}$ under pressure; the state of today's
theories \cite{bourbon:iscom-03} still ignores the challenges from
this anion ordering.

We conclude with the overview of challenges for future studies:\\
hidden existence of charge ordering and ferroelectricity in the metallic \textit{Se} subfamily;\\
optical identification of gaps and soft modes;\\
physics of solitons via conductivity, optics, NMR;\\
ferroelectric hysteresis, relaxation, domain walls;\\
more exploration of anion ordering and other structural information.

We finish to say that new events call for a substantial revision of the contemporary
picture of the most intriguing family of organic metals and its neighbors, and for
further efforts to integrate various approaches to their studies.

 \textbf{Acknowledgments.} \newline Author acknowledges collaboration
with P.~Monceau and F.~Nad, discussions with S.~Brown, H.~Fukuyama, J.-P.~Pouget and
S.~Ravy, comments from R.~Ramazashvily and S.~Teber.

\appendix

\section{Earnshow instability. Ion in the cage.\label{sec:earnshow}}

Empirically we see a systematic difference between the usual $\vec{q}\neq 0$ anion
ordering and the $\vec{q}=0$ ferroelectric transitions. The first ones are always
observed for non-centrosymmetric anions, so that the orientational ordering was supposed
to be a principle mechanism \cite{bruinsma:83}, with positional displacements being its
consequences. The accent on the orientational ordering was probably one of reasons, why
the hidden $\vec{q}=0$ transitions were not understood initially. Oppositely to
$\bf{q}\neq 0$ AOs, the $\vec{q}=0$ ones are mostly observed in systems with the
non-centrosymmetric anions. Then we should think about a universal mechanism related
only to the positional instability. This is apparently the case of the Earnshow theorem
\cite{earnshow}, which states that a classical crystal with only Coulomb interactions is
never stable. Our compounds may be regarded as such, as long as the cavity for the ion
is large enough and they are allowed to move to minimize the Coulomb energy of the
charge transfer. The Earnshow instability will probably develop itself as a displacement
$\textbf{u}$ towards one of the two closest molecules along the diagonal connecting the
nearest molecules on neighboring stacks, Fig. \ref{fig:tmttf}. Then in all cases, both
centrosymmetric and non-centrosymmetric, there is a double well potential (or the
potential is flattened \cite{pouget:05} at least) for either displacements or
orientations, or for both. It can also happen that for non-centrosymmetric ions, with
orientation as another degree of freedom, new directions of $\vec{u}$ are enforced. Then
the former ones may be abandoned, what happens probably in the relaxed
$\mathrm{TMTSF)_2ClO_4}$\footnote{In this sense, the traditional interpretation of
specifics of small anions as the "negative chemical pressure" should be revised. A small
ion is allowed to realize larger varieties of its displacive instability
\cite{pouget:05}, one of them fortunately gives rise to the $\vec q_3$ structure
favorable to superconductivity.}. Otherwise, the former displacive minima can be
preserved which opens the possibility for a sequence of transitions of $T_{\mathrm{CO}}$
and $T_{AO}$ types, as it was observed in case of $\mathrm{(TMTTF)_2ReO_4}$.

We suggest an elementary illustration of the purely ionic instability. Consider a single
ion in the cage formed by four oppositely charged molecules from two neighboring stacks.
Let $a$ is the intramolecular spacing (stack period) and $h$ is the distance from the
ion to the stack. If we allow for the ion displacements $\delta a$ along the stack, then
the energy of Coulomb interactions will change as

\[
\delta W\approx\frac{\left(h^{2}-a^{2}/2\right)} {\left(h^{2}+a^{2}/4\right)^{5/2}}
(\delta a)^2
\]
The system is unstable with respect to the longitudinal displacement $\delta a$  if
$2h^{2}<a^{2}$. Otherwise, if $2h^{2}>a^{2}$, it is unstable with respect to the
transverse displacement $\delta h$ . These two cases correspond to instabilities
observed in $\mathrm{(TMTTF)_2SCN}$ and $\mathrm{(TMTSF)_2ClO_4}$.

The quantitative criterium discriminating the types of instabilities may change if we
improve calculations by taking into account the actual charge distribution over the
large molecule, and their whole array. But the qualitative statement on the instability,
the Earnshow theorem, can be violated only by quantum mechanical effects like the
orientational energy of methyl end groups forming the cage. Altogether, the double well
potential for the ion will be formed.

These expectations give a sound interpretation of the very strong isotope effect upon
deuteration \cite{nad:06,deuter}. The $T_{\mathrm{CO}}$ enhancement can be caused
\cite{pouget:05} by the shortening of the (C-D) bonds, in comparison with the (C-H)
ones, in methyl groups, thus widening the ion's cage and reducing its stability.

\section{Permittivity sources. \label{sec:epsilon}}
\subsection{Estimations for the ionic contribution to $\varepsilon $.
\label{sec:epsilon-ions}}
 A purely ionic contribution near the $\vec{q}=0$ instability
is expected to be
\[
\varepsilon_{i}=4\pi \frac{e^{2}}{\Omega}\frac{\partial u}{\partial eE}\sim 4\pi
\frac{e^{2}}{a}\left(\frac{u_{0}}{a}\right)^{2}\frac{1}{T_{\mathrm{CO}}}
 \frac{T_{\mathrm{CO}}}{T-T_{\mathrm{CO}}}\sim
10^{1}\frac{T_{\mathrm{CO}}}{T-T_{\mathrm{CO}}}
\]
where $\Omega $ is the unit cell volume, $u_{0}\sim 0.1a$ is a guess for the equilibrium
displacement at low $T$, and also we have estimated $\partial u/\partial eE\sim
u_{0}^{2}/T_{\mathrm{CO}}$. We see that the anomaly may develop upon the background
value of only $10^{1}$ which is 3 orders of magnitude below the experimental scale of
 $\varepsilon \approx 2.5\cdot
10^{4}T_{\mathrm{CO}}/|T-T_{\mathrm{CO}}|$ \cite{monceau:01}.

\subsection{Phase instability. \label{sec:e-i}}
Consider the interference of ionic displacements and the charge ordering. Suppose the
stability of the anionic system is controlled by a parameter $K$ such that $K=0$ would
correspond to the instability due to short range interactions of charges calculated
without the contribution from electronic correlations. For homogeneous deformations and
in presence of an external electric field $E$, the energy per stack reads
\begin{equation}
\frac{1}{\pi}E\varphi -U_{b}\cos 2\varphi -U_{s}\sin 2\varphi +\frac{K}{2} U_{s}^{2}
\label{E-cos}
\end{equation}
For shortness we do not distinguish bare and renormalized values. Expanding in small
$\varphi$, the minimization yields
\[
U_{s}=\frac{E/2\pi}{U_{b}K-1}\,,\;
 \varphi=-\frac{E/\pi}{U_{b}-K^{-1}}\,,\;
 \varepsilon=\frac{4e\varphi}{sE}=\frac{4e^2/\pi s}{U_b-K^{-1}}
\]

We see that the transition of a joint electron-ion instability takes place at
$K=U_b^{-1}$, well above the point $K=0$ of a purely ionic one.

To keep nonlinearity at hand and to access the new ground state below the instability,
we should exclude from (\ref{E-cos}) only $U_{s}$ to arrive at the energy (for $E=0$)
\[
F(\varphi)=-U_{b}\cos2\varphi+\frac{1}{2K}\cos^{2}2\varphi \,+\, const
\]
At $U_{b}>K^{-1}$ there is only one minima at $\varphi =0$ but at $U_{b}=K^{-1}$ this
point becomes unstable which originates the anomaly in $\varepsilon$. At $U_{b}<K^{-1}$
the two new minima appear at $\varphi =\pm \alpha $ (first closely, with
$\alpha\approx\sqrt{(1-KU_b)/2}$ ) which makes the chain to be polarized.

\section{Competing philosophies for organic conductors.  \label{sec:philosophies}}
Since long time, there are two opposing philosophies for interpretations of these
magnificent materials:
\newline
{\bf\it A. Generic picture} summarized in \cite{jerome:99} implies that the sequence of
electronic phases follows a smooth variation of basic parameters reducible to the
effective pressure, see Fig.  5 in \cite{jerome:99}. The majority of compounds with
non-centrosymmetric anions were abandoned, presumably their AOs were thought to exhort
ill defined or undesirable complications.

The advantages are appealing:
\newline a) Concentration on simplest examples avoiding structural effects;
\newline b) Generality in a common frame of strongly correlated
systems driven mostly by basic parameters - interactions versus the bandwidth;
\newline c) Extensive use of experiments under pressure.

But there is also another side of the coin:
\newline 1) Concentration on only simplest examples avoiding
the rich information \cite{anions} on correlation of electronic and structural
properties;
\newline
2) Necessity to introduce the case of the non-centrosymmetric anion $\mathrm{ClO_4}$ to
demonstrate the appearance of the superconductivity without pressure.\footnote{The logic
of the "effective pressure" demands to show for this compound only the quenched phase
with the SDW state, rather than the relaxed phase where the superconductivity appears
only after the particular structural transition of the AO.};
\newline 3) Accent upon pressure as a universal parameter;
\newline 4) Ignoring the \lq{structureless}\rq\, transition',
 which are typical just for these selected
compounds with centrosymmetric anions; hence loosing the dominating effect of the CO/FE.

{\bf\it B. Specific picture} developed in \cite{BY:phys-let,BY:86,BY:jetp,zagreb}
suggested the synergy of structural and electronic phase transitions with the accent
upon compounds with AOs. It extends naturally to new observations on charge ordering and
ferroelectricity. Its main statements are the following (see \cite{BY:phys-let} and Ch.6
of \cite{BY:jetp} for applications):
\newline
a) Displacive, rather than orientational, mechanisms are driving the AOs (the Earnshow
instability of separated charges);
\newline
b) Each fine structural change exerts a symmetrically defined effect which triggers a
particular electronic state;
\newline
c) $1D$ "g-ological" phase diagram results in $2D$, $3D$ phase transitions only when it
is endorsed by appropriate symmetry lowering effects;
\newline
d) Main proof for the 1D physics of the Mott transitions is given by the $\vec{q}_{2}$
structure of the $\mathrm{(TMTTF)_{2}SCN}$. (Today it is seen as the AFE case of the
CO.)
\newline
e) Superconductivity appears only if the system is drawn away from the half filling thus
avoiding the Mott insulator state. It happens in the relaxed phase of the
$\mathrm{(TMTSF)_{2}ClO_{4}}$ thanks to the unique $\vec{q}_{3}$ type of the anion
ordering leading to inequivalence (CD in today's terms) of chains%
\footnote{This is a purely defined case of what today is called the "internal doping".
Estimations for its  magnitude, i.e. the potential $\pm W$ of the interchain charge
disproportionation, range from moderate $W\approx 50K$ \cite{haddad,lebed:06} to high
$W\approx 250K$ \cite{radic} and even higher \cite{BY:86} values. The uncertainty  comes
from different interpretation of the fast oscillations.}

But there are difficulties of this picture as well.
\newline 1.
In applications to $Se$ compounds there are common problems of any quasi 1D approach
confronting the success of the band picture for the FISDWs (and vice versa!).
\newline 2. There are cases of the superconducting state without observation of the particular
$\vec{q}_{3}$ type of the AO.%
\footnote{Nevertheless, the recent views on independent AOs allows to suggest that the
$\vec{q}_{3}$ structure is still there, at least in local or dynamic form without the
long range order. Low $T$ structural studies under pressure, of particularly
$\mathrm{(TMTSF)_{2}PF_{6}}$, are required.}
\newline 3. Missing opportunity to view the anion ordering transition in
$\mathrm{(TMTTF)_2SCN}$ case as the AFE charge ordering.
\newline 4. Unawareness of the later discoverd structureless transitions.

\section{History Excursions. \label{sec:history}}
Correlation between electronic phases and fine structural effects of
the AOs has been noticed \cite {zagreb,BY:phys-let,BY:jetp} and
exploited in details \cite{BY:phys-let,BY:jetp} long time ago.
Experiments generally prove this correlation but also demonstrated
deviations, see \cite{anions}, from the unique correspondence
suggested at the earlier stage. The discrepancies are related to the
third variable ingredient: the electronic dispersion in the
interchain direction. The recent events call again for a unifying
picture of electronic and structural effects which returns us to
suggestions already made two decades ago. Below we quote from ill
known publications written in early-mid 80', whose views become
relevant nowadays.

\begin{center}
\textbf{Extracts from \cite{BY:phys-let}.} For more details and applications see Ch.6 in
\cite{BY:jetp}.
\end{center}

\noindent... This theory permits us to suggest a general model for the phase diagram of
the Bechgaard salts in a way that the variation of electronic states is mainly
determined by the crystal symmetry changes.

\noindent... 1D divergent susceptibilities give rise to observable phenomena only if the
pair coherence is preserved in the course of the interchain tunneling. In the gapless
regime it can be maintained by proper interchain electronic phase shifts which can
appear due to some symmetry changes.

\noindent... Experimental data show us the following correlation between the anionic
structure (characterized by the wave vector $\vec{q}$) and
the state of the electronic system.%
\footnote{Additional notations here: PD - phase diagram, MI - magnetic/Mott insulator.}

\begin{enumerate}
\item Unperturbed structure. Bonds are dimerized. PD:
M$\rightarrow$MI$\rightarrow$ SDW.
\\The last two phases are clearly separated only in TMTTF subfamily.

\item $\vec{q}_{2}=(0,1/2,1/2)$. The molecules are not equivalent. PD:
M$\rightarrow$MI$\rightarrow$SDW (or CDW  = Spin-Peierls). $T_{\mathrm{MI}}$\ and
$T_{\mathrm{SDW}}$ are well separated \\
($\mathrm{X=SCN}$: $T_{\mathrm{MI}}=160K$, while $T_{\mathrm{SDW}}=7K$)

\item $\vec{q}_{3}=(0,1/2,0)$. The neighboring stacks are not equivalent. \\
PD: M$\rightarrow$SC$\rightarrow$FISDW.

\item $\vec{q}_{4}=(1/2,1/2,1/2)$. The tetramerization. \\
PD: M$\rightarrow$I transition being driven externally by the AO.
\end{enumerate}

\noindent ... The rare case 2. helps us to fix the model for the whole family: a
strongly correlated $1D$\ state with the separation of charge- and spin degrees of
freedom. The typical case 1. qualitatively corresponds to the same model while the
separation is less pronounced and interpretation may be controversial.%
\footnote{Fortunately it is unambiguous today, two decades later.}

\noindent  The most important for appearance of the SC is the case 3.: the alternating
potentials lead to some redistribution of the charge between the two types of stacks,
hence the system is driven from the two fold commensurability which removes the Umklapp
scattering, destroys the Mott-Hubbard effect and stabilizes the conducting state
down to lower temperatures where the SC can appear.%
\footnote{This is a clear case of the "internal doping" in today's terminology.}

\begin{center}
\textbf{\bigskip Extracts from \cite{zagreb}.}
\end{center}

Here are some extracts from \cite{zagreb}\footnote{\cite{zagreb} was probably the first
theoretical work performed in response to the discovery of the organic superconductivity
within first weeks. Hence it speaks only about the case of X=PF6: the zero pressure
superconductor X=ClO4 was not discovered yet.}, which itself was an extension of earlier
observations on effects of counterions in charge transfer CDWs of the KCP type
\cite{kcp}. A better known later publication is \cite{emery:82}.

\noindent ...We propose an alternative explanation of
$\mathrm{(TMTSF)_{2}PF_{6}}$, based on the fact that this material
possesses a weak dimerization gap $\Delta$. This gap is due to the
environment of the given chain, which, unlike the chain itself, does
not posses a screw symmetry along the chain axis. Without the effect
of the environment the band is quarter-filled. The environment
($PF_{6}$, etc.) opens a small gap $\Delta$\ in the middle of this
band, which therefore becomes half-filled. Hence also small are the
corresponding constant for the Umklapp scattering: $g_{3}\sim
g_{1}\Delta /E_{F}$. The effect of $g_{3}$ appears only below
sufficiently low temperature $T_{3}\sim
E_{F}g^{1/2}(g_{3}/g)^{1/g}$,
 $g=2g_{2}-g_{1}$.

\noindent ...Assuming the pressure suppresses $g_{3}$\ and with it $T_{3}$, the
Josephson coupling $J$ of superconducting fluctuation will finally overcome the Umklapp
scattering. This interpretation explains the observations in
$\mathrm{(TMTSF)_{2}PF_{6}}$ as a result of competition of the two small (off-chain)
parameters, $g_{3}$\ and $J$, rather than as a result of the accidental cancelation of
the large coupling constants $2g_{2}$ and $g_{1}$ (D. J´erome and H. Schulz, Adv. in
Physics {\bf 31}, 299 (1982)).

\noindent... In this way there appears a region in the phase diagram where the
superconductivity exists in absence of $g_{3}$, but where the CDW is introduced by
$g_{3}$.

\noindent... A closer examination of the model shows that it is the triplet%
\footnote{This proposal, apparent from already existing by that old time theories - see
e.g. \cite{solyom}, is back to the agenda today, 25 years later.} superconductivity.

\end{document}